%% file: magnetostatics-paper.tex
\documentclass[aps, pra, reprint, english]{revtex4-1}
\usepackage{mathtools}
\usepackage{amsmath}
\usepackage{amssymb}
\usepackage{cancel}
\usepackage{graphicx}
\usepackage{physics}
\usepackage{siunitx}
\usepackage{cleveref}
\usepackage{placeins}

\usepackage{standalone}
\usepackage{tikz}
\usepackage{xifthen}

\newmuskip\pFqmuskip

\newcommand*\pFq[6][8]{%
  \begingroup 
  \pFqmuskip=#1mu\relax
  \mathcode`\,=\string"8000
  \begingroup\lccode`\~=`\,
  \lowercase{\endgroup\let~}\pFqcomma
  {}_{#2}F_{#3}{\left(\genfrac..{0pt}{}{#4}{#5};#6\right)}%
  \endgroup
}
\newcommand{\pFqcomma}{\mskip\pFqmuskip}
\usepackage{titlecaps}
\Addlcwords{a, an, the, at, by, for, in, of, on, to, up, and, as, but, or, nor}
\newcommand\titledcaption[2][]{\caption[#1]{\ifthenelse{\isempty{#1}}{}{\titlecap{#1} ---} #2}}
\usepackage{babel}
\begin{document}

\title{Searching for Magnetostatic Modes in Spin-Polarized Atomic Hydrogen}
\author{L. Lehtonen}
\email{laanle@utu.fi}
\author{O. Vainio}
\author{J. Ahokas}
\author{J. J{\"a}rvinen}
\author{S. Sheludyakov}
\altaffiliation[Current affiliation:]{Institute for Quantum Science and Engineering,
Department of Physics and Astronomy, Texas A\&M University, College Station, TX, 77843, USA}
\author{K.-A. Suominen}
\author{S. Vasiliev}
\affiliation {Department of Physics and Astronomy, University of Turku, 20014 Turku, Finland}

\begin{abstract}
We consider a possibility of the magnetostatic type spin waves driven by a long-range magnetic dipole interactions, to account for the peaks in the ESR spectra observed in our previous work \cite{long-paper}. The Walker equation for magnetostatic modes is solved for a cylinder of atomic hydrogen, first in a uniform magnetic field and second in a linearly decreasing magnetic field. The frequency behavior of the solved modes with length of the cylinder and density of the gas is compared to experimental data. We conclude that magnetostatic modes are unlikely to account for the observed modulations of ESR spectra. 
\end{abstract}
\maketitle
\section{Introduction}
A perturbation of the local spin order in a magnetized medium may propagate over a long distance in the form of a wave. Such behavior was predicted by F. Bloch in 1929 \cite{Bloch1930} and since then has been intensively studied in different materials including ferromagnets and magnetic insulators \cite{Gurevich96, spinwaves}, liquid $^{3}\mathrm{He}$ \cite{He3BEC} and quantum gases \cite{Lee1987, guiding-and-trapping}. The large variety of linear and non-linear spin-wave phenomena  boosted interest into theoretical studies of their fundamental properties and stimulated development of numerous applications including microwave devices, telecommunication systems, radars \cite{Owens1985,Adam1988,Ustinov2010}, spintronics \cite{Schneider2008,Vogt2014,Khitun_2010,Chumak2015}, and quantum information processing \cite{Chumak2014,Khitun2012}.  Being bosons magnons can form a Bose-Einstein condensate, which was demonstrated in liquid $^{3}\mathrm{He}$ \cite{He3BEC}, cold gas of atomic hydrogen \cite{magnon-bec} and in ferromagnets at room temperature \cite{Demokritov2006}.

Spin interactions fall roughly into two categories: short-range interactions
like the nearest neighbor exchange interaction, and long-range interactions
such as the dipolar interaction between the magnetic dipole momenta
of the spins via Maxwell's laws. In spin-polarized atomic hydrogen, during collisions
the exchange interaction gives rise to the Identical Spin Rotation
Effect (ISRE) when the gas is in the quantum gas regime, i.e. non-degenerate
but the thermal de Broglie wavelength exceeds the typical size of
the atoms \cite{Bashkin_1981,Meyerovich1989,Lhuillier_1982_I}. The ISRE leads to a propagation of the spin perturbation in the form of spin waves. In our recent work we found a large variety of the spin wave modes dependent on density of the gas, magnetic field gradients, and geometry of the sample \cite{guiding-and-trapping, magnon-bec, long-paper}. We demonstrated a possibility of trapping and guiding \cite{guiding-and-trapping}
ISRE spin waves and compiled an argument for the existence of a Bose-Einstein
condensate of magnons in atomic hydrogen \cite{magnon-bec}. However some series of peaks in the ESR spectra could not be explained by the ISRE and its origin remains unclear. In this work, we turn our attention to another possible origin of spin waves, the long-range dipolar interactions and analyze a possibility for the magnetostatic spin wave modes in a high density gas. Atomic hydrogen at high densities studied in our experiments represents a unique system where these phenomena could be observed. In typical experiments with all other alkali vapours   the range of accessible densities is lower by several orders of magnitude, and effective magnetic moments of atoms are much smaller. 

Magnetostatic waves arise from the dipolar interaction between
spins in a magnetized sample. They were first observed by Griffiths \cite{GRIFFITHS1946}
and explained by Kittel \cite{Kittel1947}. A more general theory was
laid out by Walker \cite{Walker1957} who derived the Walker equation
for magnetostatic modes in a uniform field and solved it for ellipsoids.
The Walker equation has also been solved for some other geometries,
such as the infinite cylinder and infinite films, but generally the
nonuniformity of the demagnetizing field the restricts the applicability
of the Walker equation. A generalization of the Walker equation \cite{Arias2015}
overcomes this limitation, but is more difficult to solve. 

Following the features of the experimental observation of the electron spin waves, we model the
magnetostatic problem in a finite cylinder of variable length. First we just extend the results of Joseph and Schl{\"o}mann for
the infinite cylinder \cite{cylinder}, in this case for a finite cylinder. Then, a small linear gradient field is added to the problem and
an approximate solution to the Walker equation is found by simplifying boundary conditions. 

\section{Experimental observations}
\begin{figure}
\begin{centering}
\includegraphics[width=0.45\textwidth]{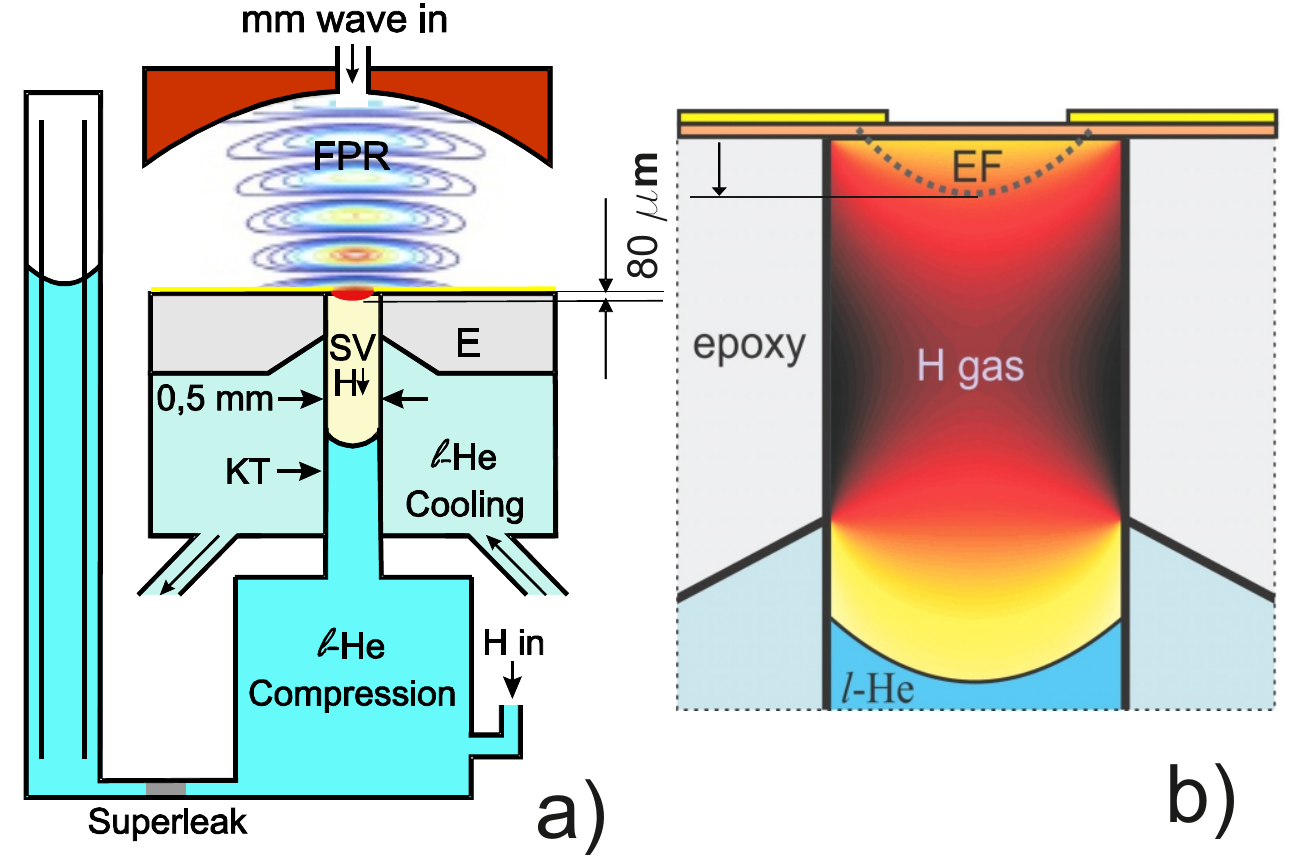}
\end{centering}
\titledcaption{Schematics of the Sample Cell Used for Compression of Atomic Hydrogen}
\label{Cell}
\end{figure}

Experiments are performed with a gas of atomic hydrogen in a strong magnetic field of \SI{4.6}{\tesla} in a temperature range of \SIrange{300}{500}{\milli \kelvin}. The gas is compressed to high densities up to $\sim  \SI{10e18}{\cm^{-3}}$ using the piston of liquid helium in a U-tube like geometry, where the compression is driven by the fountain effect of superfluid $^4$He \cite{long-paper}. Compression is performed by raising the helium level in a thin-walled polyimide tube of \SI{0.5}{\mm} diameter to a variable height ranging between \SIrange[range-phrase =~and~]{0.5}{2}{\mm} (see \cref{Cell}). This is done by driving down the level in the other arm of the U-tube system by decreasing the temperature and fountain pressure in the latter. The compression is stopped after reaching a certain height. Highest densities are reached for largest height and volume changes, i.e. stopping the ramp at smallest heights. After stopping the compression, the sample evolves with the recombination processes decreasing number of atoms, density and pressure of the sample. Decrease of pressure of the gas leads to further reduction of its height. Therefore, the height and density appear to be bound parameters of the experiment. In order to obtain different values of the gas density for the same height, we run a series of compressions with different values of the final height at compression stage.

For diagnostics of the compressed gas we use Electron Spin Resonance (ESR) technique \cite{RSI} at \SI{130}{\GHz}. Excitation of the gas is performed via the evanescent microwave field, penetrating from the high Q Fabry-Perot cavity into the compression cylinder via a \SI{0.5}{\milli \meter} coupling orifice.  The characteristic size of the microwave field region is $\sim$ \SI{80}{\micro \meter}, substantially smaller than the height of the cylinder with the gas. Despite of a special care taken to avoid any magnetic impurities in the sample cell, we found that even essentially "nonmagnetic" materials like the Stycast 1266 epoxy may influence the magnetic field homogeneity at the levels sufficient to influence the magnetic resonance of the compressed gas. The presence of the weakly diamagnetic epoxy created $\sim$ 0.3 G local field maximum near the walls of the cylinder. Since for the ISRE magnons the field inhomogeneities are equivalent to the change of the potential, such field profile allowed formation of the standing modes of the ISRE magnons in the regions of magnetic field maximum \cite{magnon-bec}.

In order to be able to change the static magnetic field homogeneity we installed a pair of extra coils in the anti-Helmholtz geometry, which could create axial gradients of magnetic field up to 30 G/cm. Using these gradient coils we found that in the case of large ($>$10 G/cm) positive gradient the ISRE magnon modes are localized near the upper end of the cylinder, in the maximum of magnetic field \cite{magnon-bec}. In the opposite case of the large negative gradient, our rf excitation launched traveling ISRE waves going towards the lower end of the cylinder down the magnetic potential \cite{guiding-and-trapping}.

\begin{figure}
\begin{centering}
\includegraphics[width=0.45\textwidth]{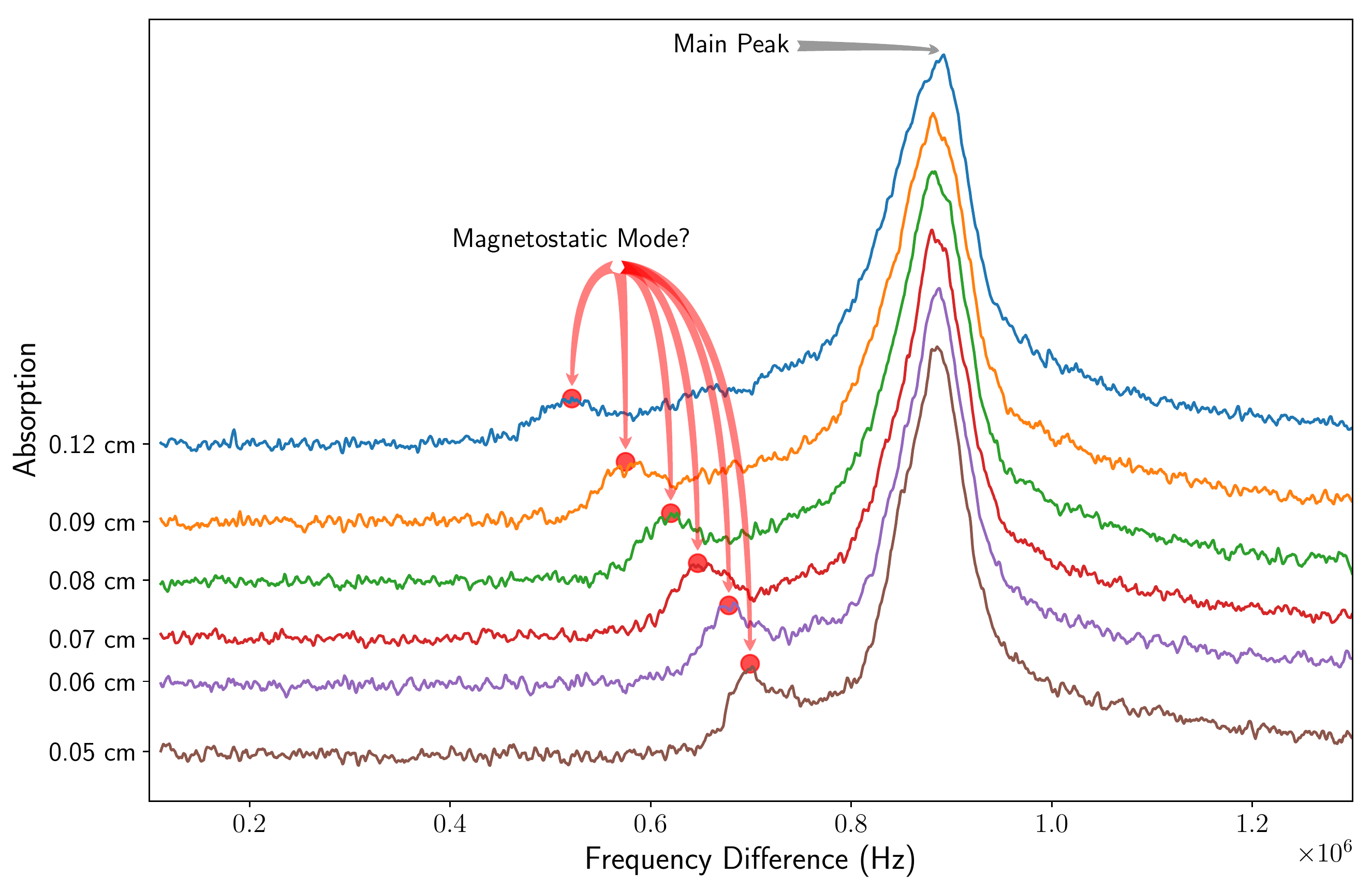}
\end{centering}
\titledcaption[ESR spectra of the compressed hydrogen gas recorded after the compression ramp.]{The density and height of the sample decrease from top to bottom spectra. The modes of possible magnetostatic origin considered in this work are outlined by arrows. They appear below the main ESR peak which is found at the Larmor frequency in the static magnetic field of our experiments.}
\label{ESR peaks}
\end{figure}

The observed peaks in the ESR spectra in large gradients as well as in the local maxima of the static field are understood and interpreted in terms of the ISRE. However, in the case of the most homogeneous static field, with the only small disturbances originating from the diamagnetic epoxy, we observed a set of the peaks in the ESR spectrum dependent of the H density and on the height of the gas column. These peaks occurred below the main ESR peak in frequency space, approaching it as the density and height of the sampled decreased during the decay of the sample (see \cref{ESR peaks}). They behave differently from the ISRE modes, which are typically concentrated in the regions of stronger magnetic field and appear above the main peak. We were not able to explain it by the ISRE effect. Motivated by this discrepancy we consider another possibility for the spin waves in magnetized media, well known in solid-state physics, namely magnetostatic waves.

\begin{figure}
\begin{centering}
\includegraphics[width=0.45\textwidth]{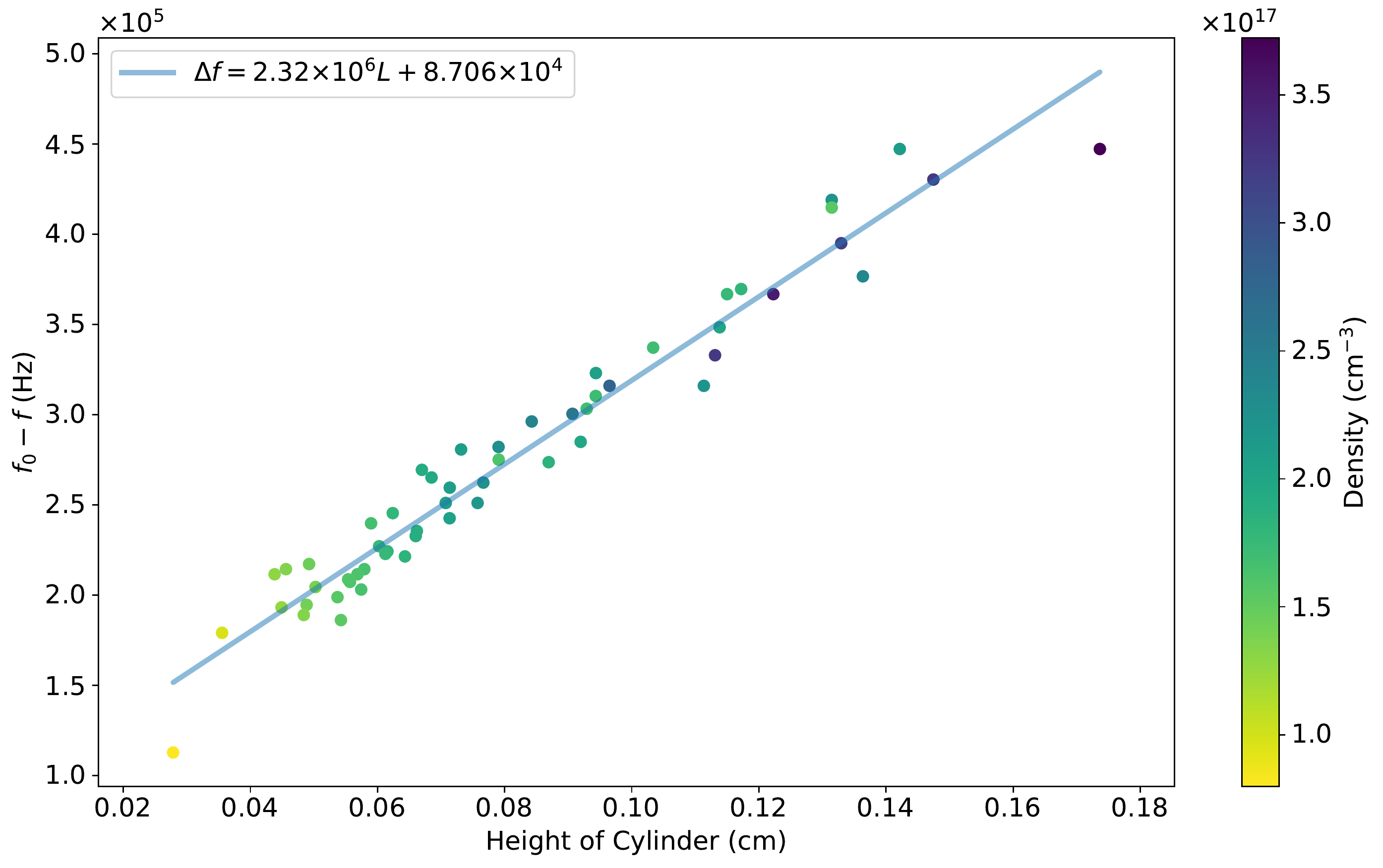}
\end{centering}
\titledcaption[Slope of frequency difference]{
The figure shows the linear behavior of the frequency displacement from the main peak when plotted against the height of cylinder. While the mode frequency decreases, their displacement from the main peak increases.}
\label{Single compression data}
\end{figure}

In \cref{Single compression data} we plot the data of the displacement of the observed spin-wave peaks as a function of the cylinder height taken from a single compression experiment. By  virtue of the compression experiment, the density of the gas decreased proportionally to the height, which is labeled by the color of the symbols (see color map on the right of the figure). Even though the plot reveals quite a clear linear dependence, such data is difficult to analyze. 

\begin{figure}
\begin{centering}
\includegraphics[width=0.45\textwidth]{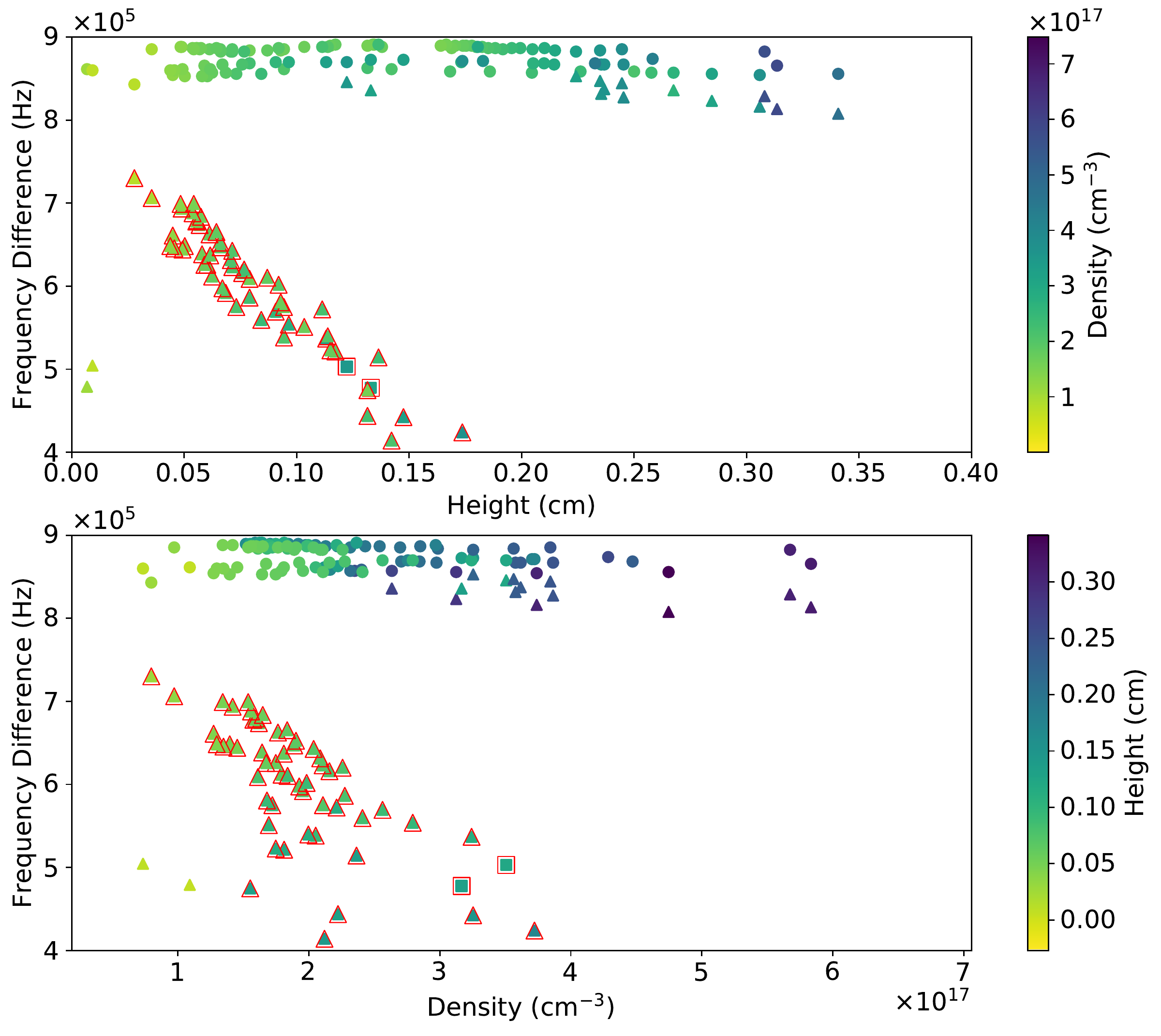}
\end{centering}
\titledcaption[Positions of main and side peaks observed in experiment]{
  The two subfigures show the same data with a different $x$-axis: height of the
  gas cylinder in the upper and density of hydrogen gas in the lower. The other variable
  is indicated by the color as shown in the colorbar, so the color encodes different data
  in the subfigures. The modes of interest are selected based on their behavior in the
  upper plot and are shown with a red outline; the same modes are outlined in the lower plot.
  The data points near the top of the figure with round markers show the position of the main peak;
  triangular markers and rectangular markers show the positions of the first and second peak counting down
  from the main peak, respectively. While occasionally the spectra had three peaks in this set of data,
  in most of them only two peaks were visible. }
\label{Data vs height and density separated}
\end{figure}

In order to separate the dependence of the experimental parameters, we performed a series of compressions with different initial height and density. This allows selection of the spectra for same density and different height, as well as the same height but variable density. The plots of the frequency displacement of the modes from the main ESR peak ($\omega_0 - \omega$) against gas column height and gas density are shown in \cref{Data vs height and density separated}. 

\section{Finite Cylinder in Uniform Magnetic Field}

\begin{figure}
\begin{centering}
\includegraphics[width=0.45\textwidth]{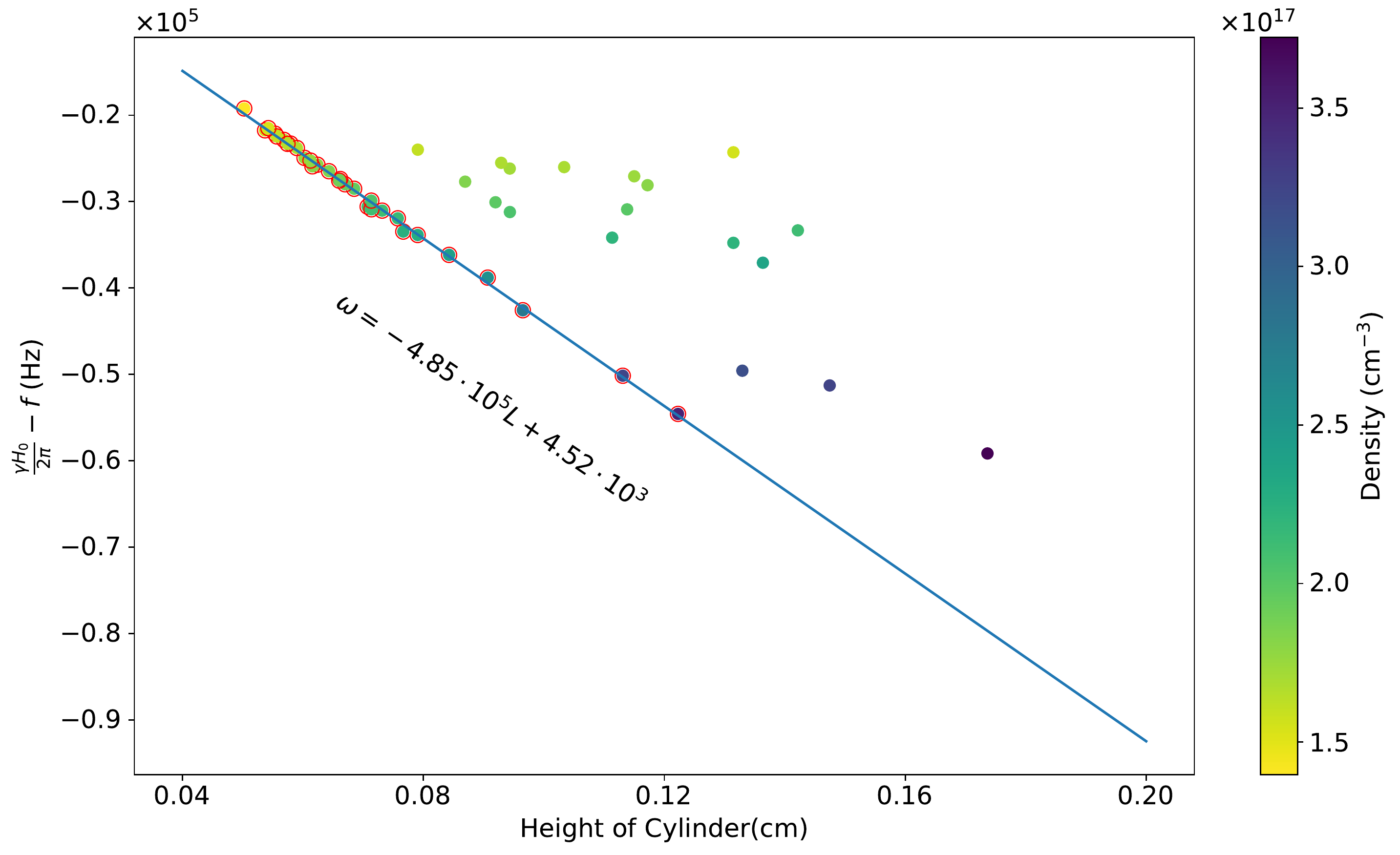}
\end{centering}
\titledcaption[Frequency displacement of pure Walker magnetostatic modes]{
The frequencies of the modes are calculated using Walker's equation
for a finite cylinder for experimental cylinder heights and gas densities.
Demagnetizing field is ignored completely and the frequency of the
main peak is assumed to be constant. The behavior of the modes is
clearly in the opposite direction compared to the experiment(~\cref{Single compression data}), and
the magnitude is off by a factor of 5. Notably, the modes seem to
exist above the main ESR peak $\omega_0=\gamma H_{0}$ and grow more distant
from it as the length of the cylinder grows (cf. \cref{ESR peaks}).}
\label{nondemag-behaviour}
\end{figure}

By dimensional analysis one would expect a slope of magnitude

\begin{align*}
\frac{\gamma_{e}\mu_{0}M_s}{L} & =\\
  = & \frac{\SI{2.8e10}{\Hz\per\tesla}
      \cdot
      \SI{1.25e-4}{\tesla\cm\per\A}}{\SI{0.1}{\cm}}\\
  \times &
  \overbrace{\SI{5e17}{\cm^{-3}}\cdot\SI{9.27e-20}{\A\cm^{2}}}^{\approx\SI{0.046}{\A\per\cm}} \\
\approx & \SI{1.6e6}{\Hz\per\cm},
\end{align*}

which is very close to the measured dependence, so an explanation
in terms of magnetostatic modes seems at least initially plausible. Magnetostatic modes or Walker modes are a specific kind
of approximate solution of the precessing magnetization equation:
\begin{equation*}
\dv{\vec{M}}{t}=\gamma(\vec{M}\times\vec{H}).
\end{equation*}
One assumes that  $\vec{H} = \vec{H_0}+\vec{h}$ and $\vec{M} = \vec{M_s}+\vec{m_{\perp}}$ to solve $\vec{m_{\perp}}$ in terms of $\vec{h}$, and then uses the Gauss' law for magnetic field to derive the Walker equation\cite{Walker1957}:
\begin{equation*}
  \mu\left[\pdv[2]{x}+\pdv[2]{y}\right]\Psi(\vec{r})+\pdv[2]{\Psi(\vec{r})}{z}=0.
\end{equation*}
The solutions in terms of Bessel functions are found in \cref{homogeneous-cylinder}; the frequencies are solved numerically from the boundary conditions \cref{uniform-characteristic-equation} for experimental cylinder lengths and gas densities. The calculated frequencies for certain modes as a function of cylinder length are shown in \cref{nondemag-behaviour}. 

It turns out that generally magnetostatic modes appear when $\omega \in [\gamma H, \gamma \sqrt{H B}]$,
however experiment tells us that the observed mode always appears below the main peak, whose frequency is
 $\omega_0=\gamma H_0$. This conflict is clear in \cref{nondemag-behaviour}:
the distance $\omega_0 -\omega$ is negative even if the magnitude of the slope is off by only a factor of 5.
It is possible to add an averaged demagnetization factor to the field (see \cref{homogeneous-cylinder}), but it only slightly changes the offset and does not significantly affect the slope.

The effect of using $\omega_B=\gamma \left(H_0+4\pi M_s(\rho)\right)$ instead of $H_0$ is to essentially flip the sign of the slope (because it would move the main peak above the modes). However, the data are best explained by the ad hoc $H_0+12 \pi M_s(\rho)$, for which the slope is  $\SI{2.34e6}{\Hz\per\cm}$. 
As a possible source of this kind of dependence we consider the uniform Kittel (magnetostatic) mode \cite{Kittel1948}, whose frequency in an infinite cylinder would be $\gamma (H +2 \pi M_s(\rho))$; unfortunately
this does not even manage to flip the sign of the slope, and the cylinder is in any case far from infinite with a $\frac{\textrm{radius}}{\textrm{length}}>0.1$.

\begin{figure}
\begin{centering}
\includegraphics[width=0.45\textwidth]{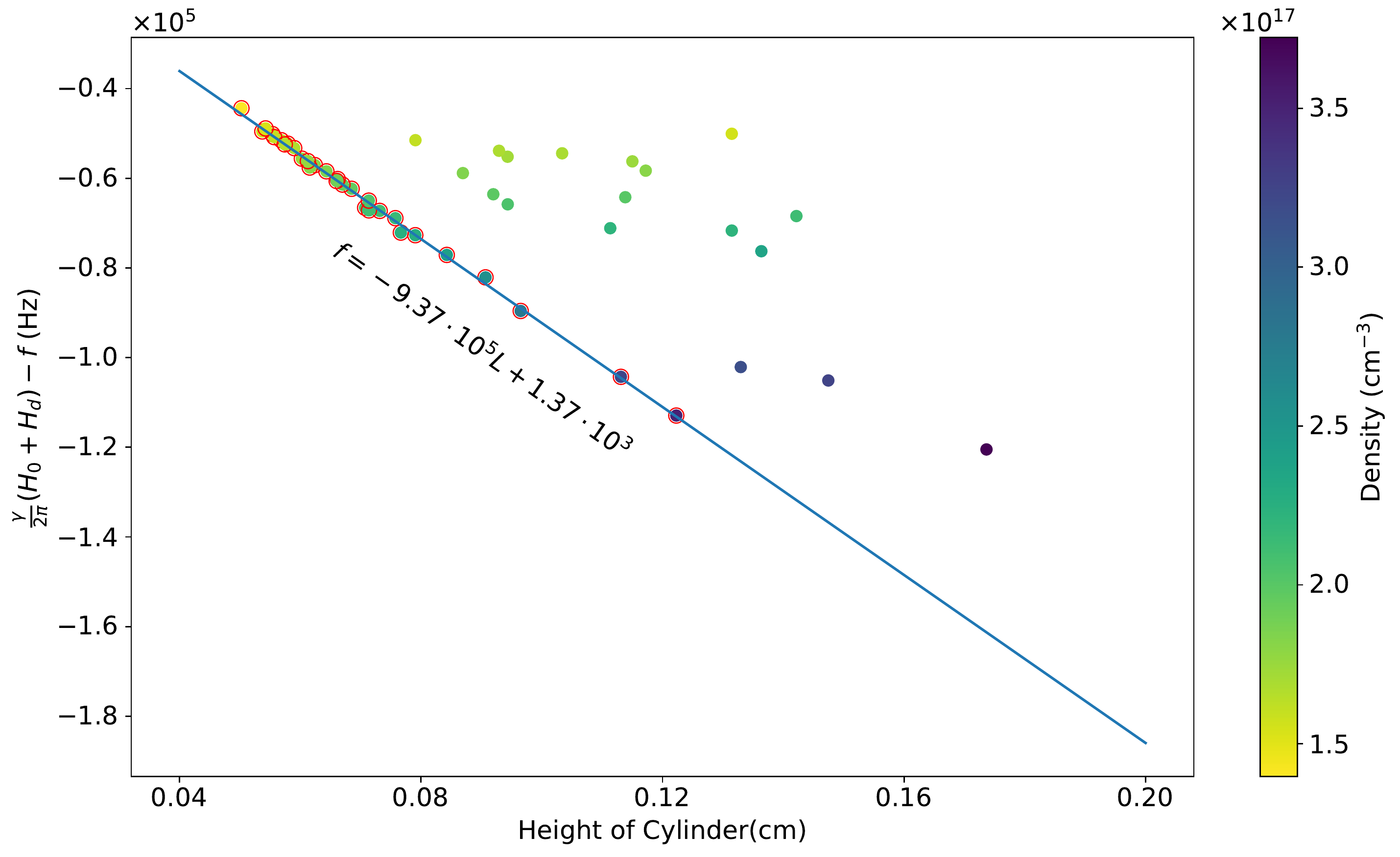}
\end{centering}
\titledcaption[Expected Behavior of Modes with Respect to the Main Peak]{
Modes calculated as in \cref{nondemag-behaviour}, but the main peak's frequency
is assumed to be $f=\frac{\gamma}{2 \pi } (H_0 + H_{d, z})$ (\cref{demag-field-cylinder-ends}). The magnitude of the slope is off by factor of 2.5, but in the wrong direction, so the result is worse than in \cref{nondemag-behaviour}.}
\label{demag-mainpeak}
\end{figure}

 Although the magnetization should not directly affect the main peak frequency, it should affect it through the demagnetizing field, however one would expect it to lower the main peak frequency (a demagnetizing field should reduce the field), making the discrepancy worse. In a finite cylinder along the $z$-axis the demagnetizing field can be solved (\cref{calculating-demagnetizing-field}); at the ends of the cylinder (where the exciting field is concentrated) it has a simple expression:
  \begin{equation}
    H_{d,z}(0,0, L) = -4\pi M_s+\frac{4\pi M_s}{2}\frac{L}{\sqrt{r_{c}^{2}+L^{2}}}. \label{demag-field-cylinder-ends}
  \end{equation}
  The result in shown in \cref{demag-mainpeak}. As expected, the shift is in the wrong direction, although the magnitude of the slope is only off by a factor or 2.5.

  As stated previously, modes were only found in range $\omega \in [\gamma H, \gamma \sqrt{H B}]$, which corresponds to the region where the parameter $\mu$ is negative. In the case of the infinite cylinder, some solutions were also found for positive $\mu$, that is at frequencies below the main peak. In the case of finite cylinders, those solutions seem to be excluded by the boundary conditions at the ends of the cylinder (see also \cref{positive-mu-absence}).

All in all, the model one chooses for the main peak significantly affects the predicted behavior,
and the most plausible models for the main peak do not succeed in predicting the observed length/density dependence.

\section{Linear Gradient in a Finite Cylinder}
In some of the experiments the calculated magnetic field inside the
gas cloud has a small linear gradient. As the height of the gas column changes, the
minimum or the maximum of the magnetic field moves, so it is not quite so obvious what happens to the modes --- especially
when the spins can only be tilted near one end of the cylindrical container: the gradient could well shift some of the modes
below the main peak. On the other hand, in some respects this gradient facilitates the treatment of the problem, as the demagnetizing field
can be counted as a part of the gradient. However generally the solution becomes more difficult due to introduction
of $z$-dependence to the $\mu$-factor ($H_0 \to H_0-C z$), which makes both the axial differential equation and boundary conditions more difficult to solve. In addition the Walker equation is slightly modified by the presence of the gradient, but this turns out to not play a major role.

\begin{figure}
\begin{centering}
\includegraphics[width=0.8\textwidth]{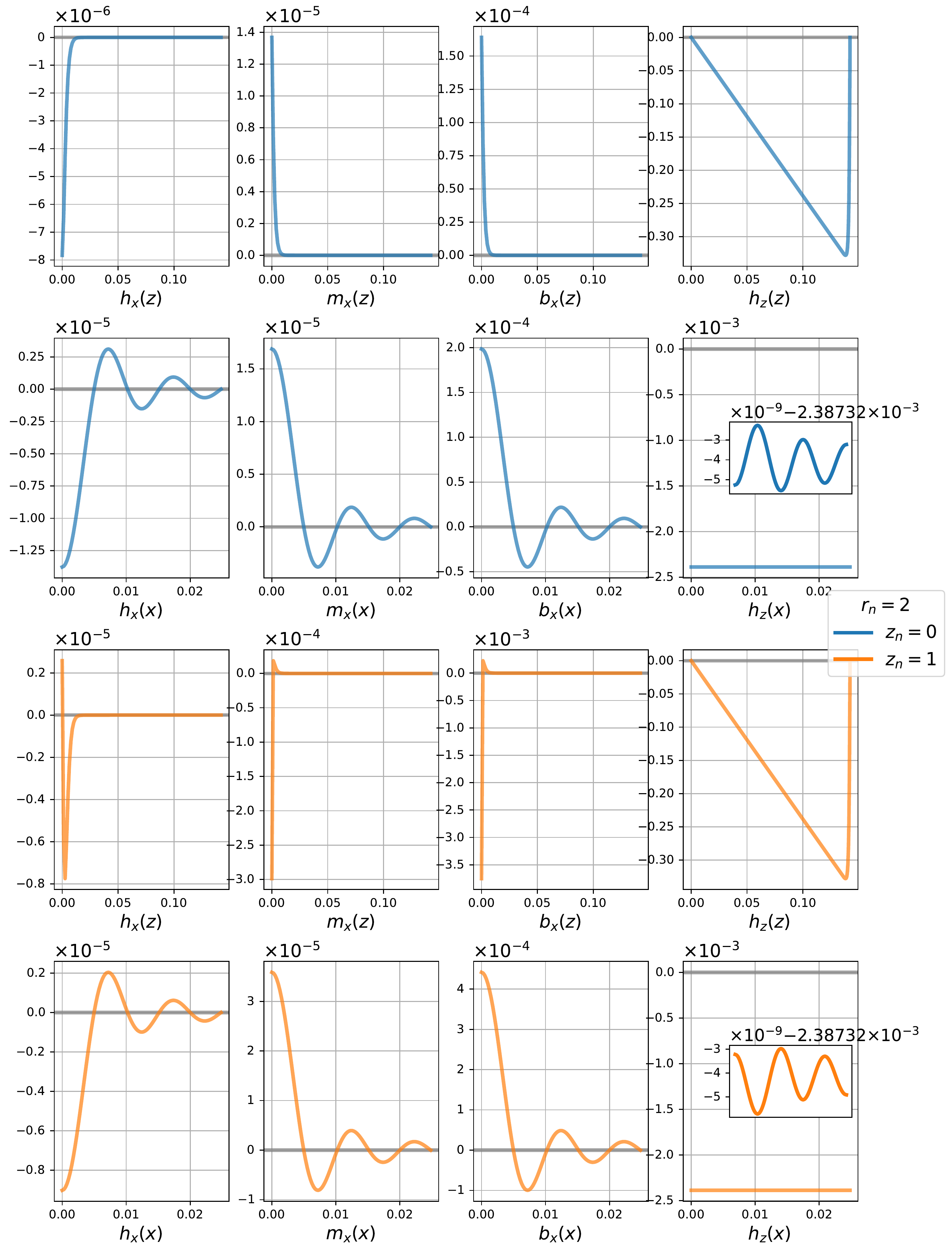}
\end{centering}
\titledcaption[Two solutions of the Walker equation in a linear gradient]{
Solutions $r_n=2, z_n = 0, 1$ with $n=\SI{2.1e17}{\cm^{-3}}$ and $L=\SI{0.142}{\cm}$. $m_z=0$ by assumption
so $h_z$ = $b_z$. The $x$-component along the $z$-axis has a clear turning point around
$z\approx 0$ for $z_n=1$, while for $z_n=0$ the fields decay more or less exponentially. Along
the $x$-axis $h_z$ is practically constant.}
\label{gradient-mode}
\end{figure}

The solution is derived in \cref{linear-gradient-solution} in terms of the confluent hypergeometric function $_1F_1$ and the Kummer $U$-function, using simplified boundary conditions where the $\vec{h}$-field vanishes at the boundary. The frequencies are again found numerically from \cref{linear-gradient-characteristic-equation}.   An example of a solution appears in \cref{gradient-mode}. Only a handful of modes were found for the experimental $C = 3.0\, \text{G/cm}$, compared to the uniform field case where countably infinite modes
exist. At a lower gradient the number of modes increased, suggesting that the gradient somehow suppresses magnetostatic modes. 

\begin{figure}
\begin{centering}
\includegraphics[width=0.45\textwidth]{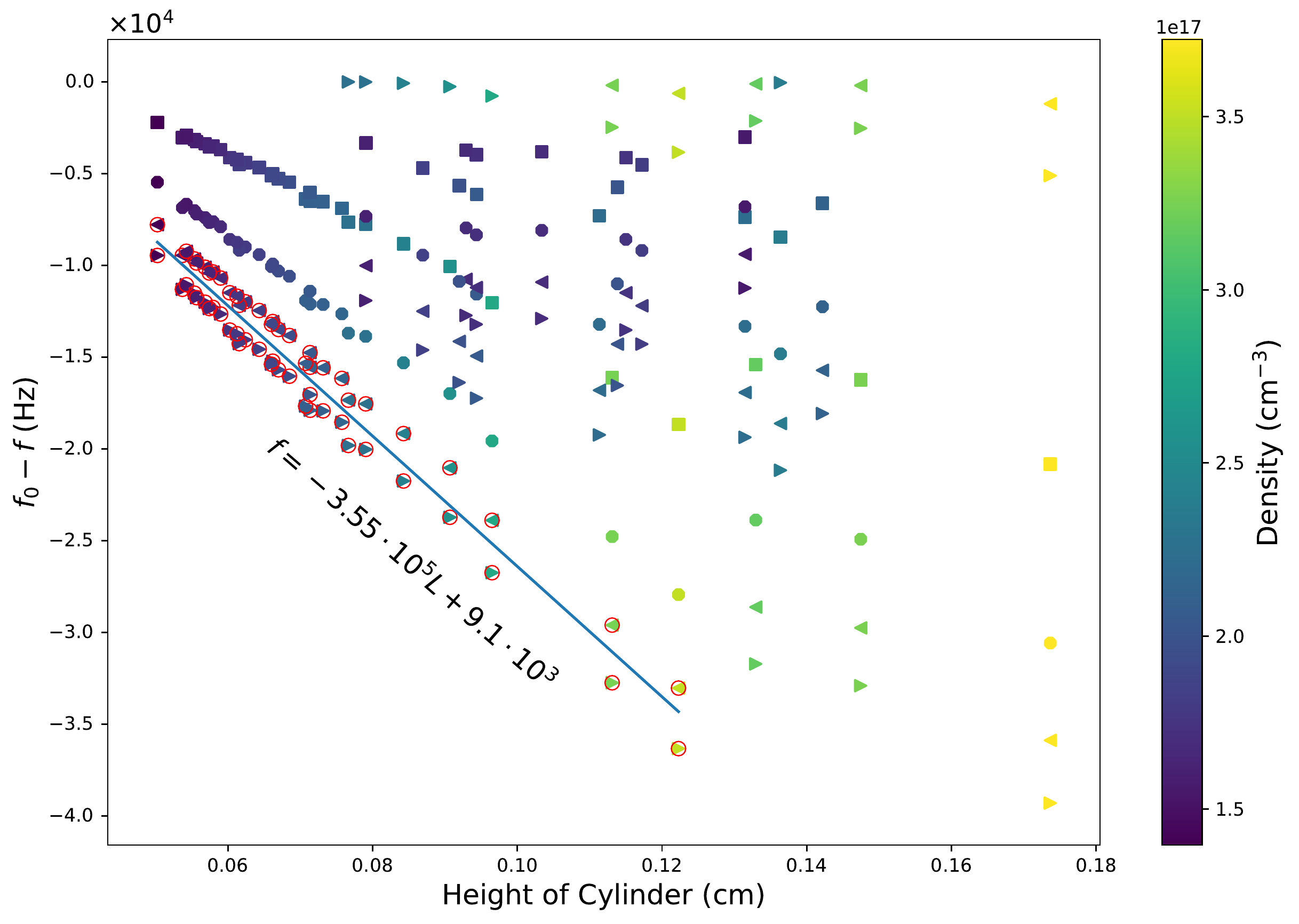}
\end{centering}
\titledcaption[Simulated Frequency Displacement of Walker Magnetostatic Modes In
Linear Gradient]{
The modes are calculated in \cref{linear-gradient-solution}, with $\alpha=6.52\cdot10^{-5}$, corresponding to $C=3.0\, \text{G/cm}$. The markers correspond to different radial mode numbers. 
The main peak is located at $\omega_0 = \gamma H_0$.
The figure shows some axial and radial modes with mode numbers around 10,
as they have the highest slopes. The slopes initially seem to increase with
$z_n$ but appear to reach some maximum value close to that of the figure.
The situation is quite similar to \cref{nondemag-behaviour}: the slope is a factor of 7 too small,
on the wrong side of the main peak, and the slope has the wrong sign.}
\label{linear-gradient-adjusted-frequency}
\end{figure}

The calculated mode frequencies are shown in \cref{linear-gradient-adjusted-frequency} as a function of height of gas cylinder for experimental heights and densities. The situation mirrors \cref{nondemag-behaviour} and appears unable to explain
the observed slope of the modes. Surprisingly no modes were found below the main peak frequency $\gamma H_0$ (the maximum field), counter to the assumption that the field gradient would shift the modes below the main peak. 
\clearpage
\onecolumngrid

\clearpage
\twocolumngrid
\clearpage
\section{Discussion}
Two models were examined as candidate explanations of the data in \cref{Single compression data} and \cref{Data vs height and density separated}: naive model that ignored the demagnetizing field, and a simplified model with a linear gradient of magnetic field $C z \vu{e}_z.$ While the magnitude of the predicted behavior is not off by more than a factor of 2-7, its direction is wrong, putting the predicted modes on the wrong side of the main peak; in this, the models are similar. This is not altogether surprising in retrospect, as the range of frequencies where the Walker equation and its generalizations are expected to have oscillatory, wave-like solutions is\cite{Arias2015}
\begin{equation}
  \label{magnetostatic-frequency-range}
 \gamma H(\vec{r}) < \omega < \gamma \sqrt{H({\vec{r}) B(\vec{r})}},
\end{equation}
where the field $\vec{H}$ is a sum of the applied field $\vec{H_0}$ and the demagnetizing field $\vec{H_d}$. That no modes were found below the main peak is somewhat surprising, as in an infinite cylinder some do exist and one could expect to see some in a linear gradient setup. While such modes may not be categorically excluded, a better accounting of the demagnetizing field and possibly the inclusion the ISRE effect (as is done in \cite{nymanthesis}) are probably required to find them. 

Although the peaks in question are unlikely to be among the found magnetostatic modes, it is interesting to examine the question of whether we should be able to see any of the found modes. In the case of uniform field, \cref{magnetostatic-frequency-range} evaluates to a frequency range of width
$\Delta \omega \sim 6-60\, \text{mG}$, which places them on top of our main peak. It is then possible that the modes are excited and modify the main peak, along with ISRE. 

\begin{figure}
\begin{centering}
\includegraphics[width=0.45\textwidth]{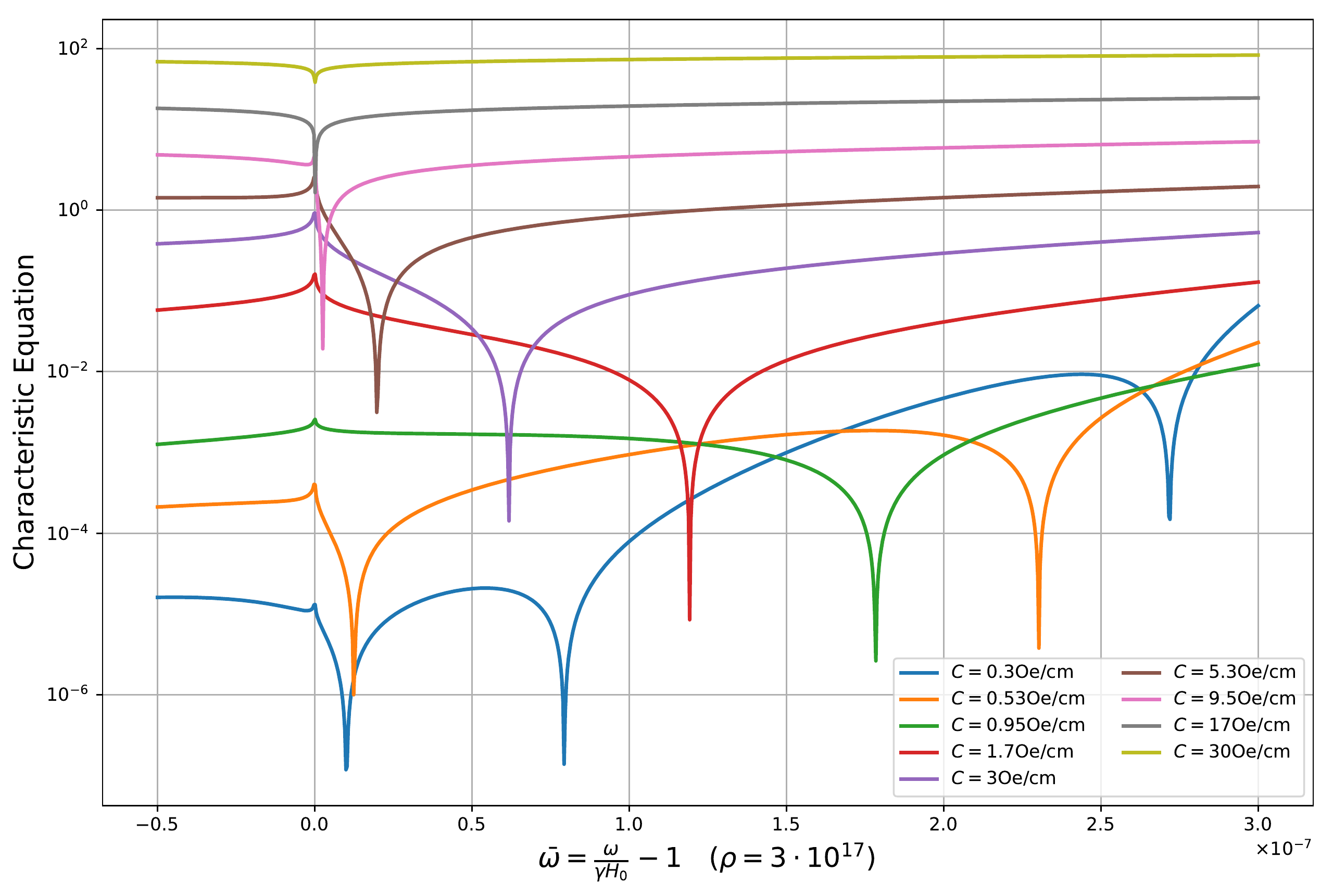}
\end{centering}
\titledcaption[Characteristic equation of Walker problem in linear gradient]{
Log-scaled magnitude of the characteristic equation as a function of scaled frequency $\bar{\omega}=\frac{\omega}{\gamma H_0}-1$.}
\label{linear-gradient-characteristic-eq-at-different-gradients}
\end{figure}

In the case of the linear gradient, the frequency range is an order of magnitude larger, $\Delta \omega \sim 70-600 \, \text{mG}$, so in certain conditions some of the modes could be separate from the main peak. On the other hand,  modes were only found above $\omega_0=\gamma H_0$, so instead of the gradient shifting the modes below the main peak it seems more like they vanished.
This is supported by the nature of the characteristic equation \cref{linear-gradient-characteristic-equation}, shown in \cref{linear-gradient-characteristic-eq-at-different-gradients}. As the gradient increases, the roots (= downward cusps) of the characteristic equation move towards $0$ ($=\omega_0$), the main peak. It is possible that not all roots were found, but if there are roots beyond $0$, they are likely complex roots. Despite some effort, none were found. In any case, it seems that magnetostatic modes only exist above the main peak, and it may  be that due to modes vanishing, the modes in the linear case  are not any more separated from the main peak than in the uniform field case.

\begin{acknowledgments}
L.L. was supported by the Vilho, Yrj{\"o} and Kalle V{\"a}is{\"a}l{\"a} Foundation of the Finnish Academy of Science and Letters during this work. This work was also supported by the Academy of Finland (grants nos. 122595 and 133682) and the Wihuri Foundation.
\end{acknowledgments}

\bibliographystyle{plain}
\bibliography{magnetostatics-paper}

\appendix

\section{Solution of the Walker Equation in a Finite Cylinder with Uniform
Magnetic Field} \label[appendix]{homogeneous-cylinder}

We seek solutions to the so-called Walker equation for magnetostatic
modes. A magnetic dipole in a magnetic field experiences a torque $\dv{\vec{J}}{t} = \vec{\mu} \times \vec{B}$.
The magnetic moment $\vec{\mu}$ is related to the angular momentum $\vec{J}$ by $\vec{\mu} = \gamma_e \vec{J}$, where
$\gamma_e$ in our case is the gyromagnetic ratio of the electron. This leads to the equation of a precessing magnetic moment: 
\begin{equation*}
\dv{\vec{\mu}}{t}=\gamma_e(\vec{\mu}\times\vec{B}). 
\end{equation*}
Within media consisting of multiple magnetic moments, one typically looks for an equation for a precessing magnetization:
\begin{equation}
\dv{\vec{M}}{t}=\gamma (\vec{M}\times\vec{H_{\text{eff}}}) \label{eq:landau-lifshitz}
\end{equation}
where in SI units $\gamma = \gamma_e \mu_0$ and $\gamma = \gamma_e$ in Gaussian units, and $H_{\text{eff}}$ is the effective
magnetic field which may have various contributions depending on the medium; in our case, only the applied field $H_0$ and the demagnetizing field $H_d$ are relevant.

The Walker equation arises as an approximate solution of this equation. First we assume that $\vec{H}$ and $\vec{M}$ have the following form: 

\begin{align}
\vec{H} & =  H_{0}\vu{e}_z+\overbrace{(h_{x},h_{y},h_{z})}^{\vec{h}}e^{i\omega t}\label{eq:hparts}\\
\vec{M} & =  M_{s}\vu{e}_z+\underbrace{(m_{x},m_{y},0)}_{\vec{m}}e^{i\omega t}.\label{eq:mparts}
\end{align}

Here $H_0$ is the applied field, and $M_s$ the saturation magnetization. We then substitute this into \cref{eq:landau-lifshitz}:

\begin{align}
  \begin{split}
 &  i\omega(m_{x},m_{y},0)e^{i\omega t} \\
 = & \gamma(m_{y}H_{0}-h_{y}M_{s},h_{x}M_{s}-m_{x}H_{0},0)e^{i\omega t}\label{eq:landau-applied}
   \end{split}\\
  \Rightarrow & \nonumber \\
4\pi m_{x}  = & \underbrace{\frac{4\pi\gamma^{2}H_{0}M_{s}}{\gamma^{2}H_{0}^{2}-\omega^{2}}}_{\kappa}h_{x}-i\underbrace{\frac{4\pi\gamma M_{s}\omega}{\gamma^{2}H_{0}^{2}-\omega^{2}}}_{\nu}h_{y}\label{eq:magn-x}\\
4\pi m_{y}  = & i\nu h_{x}+\kappa h_{y}.\label{eq:magn-y}
\end{align}

Using the relation $\vec{B}=\vec{H}+4\pi\vec{M}$ and Gauss' law for the magnetic field, we arrive to the Walker equation:

\begin{align}
  \begin{split}
0= & \nabla\cdot\vec{B}= \nabla^{2}\overbrace{\Psi}^{\vec{h}=\nabla\Psi}+4\pi\nabla\cdot\vec{m} \\
= & \nabla^{2}\Psi+\pdv{\left(\kappa h_{x}-i\nu h_{y}\right)}{x}+\pdv{\left(i\nu h_{x}+\kappa h_{y}\right)}{y}\\
= & \nabla^{2}\Psi+\kappa\pdv[2]{\Psi}{x}-\cancel{{i\nu\pdv{\Psi}{y}{x}}}+\cancel{i\nu\pdv{\Psi}{y}{x}}+\kappa\pdv[2]{\Psi}{y} \\
= & \left[\underbrace{(1+\kappa)}_{\mu}\left(\pdv[2]{x}+\pdv[2]{y}\right)+\pdv[2]{z}\right]\Psi(\vec{r}).
  \end{split}\label{eq:walker-derivation}
\end{align}
So in cylindrical coordinates the Walker equation reads

\begin{equation}
  \mu\left[ \pdv[2]{r} + \frac{1}{r}\pdv{r} + \frac{1}{r^{2}}\pdv[2]{\theta} \right]\Psi(\vec{r})
  + \pdv[2]{\Psi(\vec{r})}{z}.
  \label{eq:walker-equation}
\end{equation}

Outside the magnetized cylinder of gas, $\kappa=0$ and the equation
reduces to the Laplace equation. 

The boundary conditions are those common to Maxwell's equations: the
continuity of $\Psi$, the continuity of the tangential component
of $\vec{h}$, and the continuity of the normal component of $\vec{b}=\vec{h}+4\pi\vec{m}$
at all boundaries. In practice one needs to worry about
the continuity of $h_{z}$ across the ends and the continuity of $b_{r}$
across the sides. The latter gives the following matching condition
for radially symmetric solutions with radial number $l:$
\begin{equation}
\mu\pdv{\Psi^{\text{in}}}{r}+\frac{l\nu}{r}\Psi^{\text{in}}=\pdv{\Psi^{\text{out}}}{r}.\label{eq:radial-boundary-condition}
\end{equation}
Further, as both $\vec{h}$ and $\vec{b}$ are generated by local
fields and finite cloud of gas, they are expected to vanish as $|\vec{r}|\to\infty$,
i.e. $\Psi(\vec{r})$ tends to some constant at large distances from
the setup.
\begin{figure}[h!t]
    \centering
    \includestandalone[mode=tex, width=0.15\textwidth, keepaspectratio]{cylinder-area}
\titledcaption{Areas of the Cylinder}
\label{cylinder-areas}
\end{figure}
Solutions are only found for $\mu<0$ (\cref{cylinder-areas}):
\begin{widetext}
\begin{equation}
  \Psi(r,\theta,z)=
  \begin{cases}
A J_{l}\left(kr\right) e^{i l\theta} \left(\sqrt{-\mu}\cos(k\sqrt{-\mu}z)+\sin(k\sqrt{-\mu}z)\right) & \text{in A}\\
\begin{cases}
A \sqrt{-\mu}J_{_{l}}(k r) e^{i l\theta} e^{kz} & z<0\\
\left(\frac{1}{2}A (1-\mu)\sin(k\sqrt{-\mu}L)e^{kL}\right)J_{l}(kr)e^{i l\theta}e^{-kz} & z>L
\end{cases} & \text{in C}\\
\left(A \frac{J_{l}(kr_{c})}{K_{l}(k\sqrt{-\mu}r_{c})}\right)K_{l}(k\sqrt{-\mu}r)e^{i l\theta}\left(\sqrt{-\mu}\cos(k\sqrt{-\mu}z)+\sin(k\sqrt{-\mu}z)\right) & \text{in B}
\end{cases}
\end{equation}
\end{widetext}
with the condition
\begin{equation}
  \label{wavenumber-k}
  k=\begin{cases}
\frac{1}{\sqrt{-\mu}L}\left[\tan^{-1}\left(\frac{-2\sqrt{-\mu}}{1+\mu}\right)+n\pi\right] & \mu < -1\\
\frac{1}{\sqrt{-\mu}L}\left[\tan^{-1}\left(\frac{-2\sqrt{-\mu}}{1+\mu}\right)+(n+1)\pi\right] & \mu > -1 
\end{cases}
\end{equation}

and the characteristic equation given by boundary conditions from which the frequencies are solved:
\begin{multline}
  \label{uniform-characteristic-equation}
  J_{l}(kr_{c})K_{l}^{\prime}(k\sqrt{-\mu}r_{c})k\sqrt{-\mu}-K_{l}(k\sqrt{-\mu}r_{c}) \\
  \times \left[\mu kJ_{l}^{\prime}(kr_{c})+\frac{l\nu}{r_{c}}J_{l}(kr_{c})\right]=0.   
\end{multline}
The modes are solved numerically with a Python program using IPython, SciPy, and NumPy \cite{ipython,numpy,scipy}. 

\begin{figure}
\begin{centering}
\includegraphics[width=0.45\textwidth]{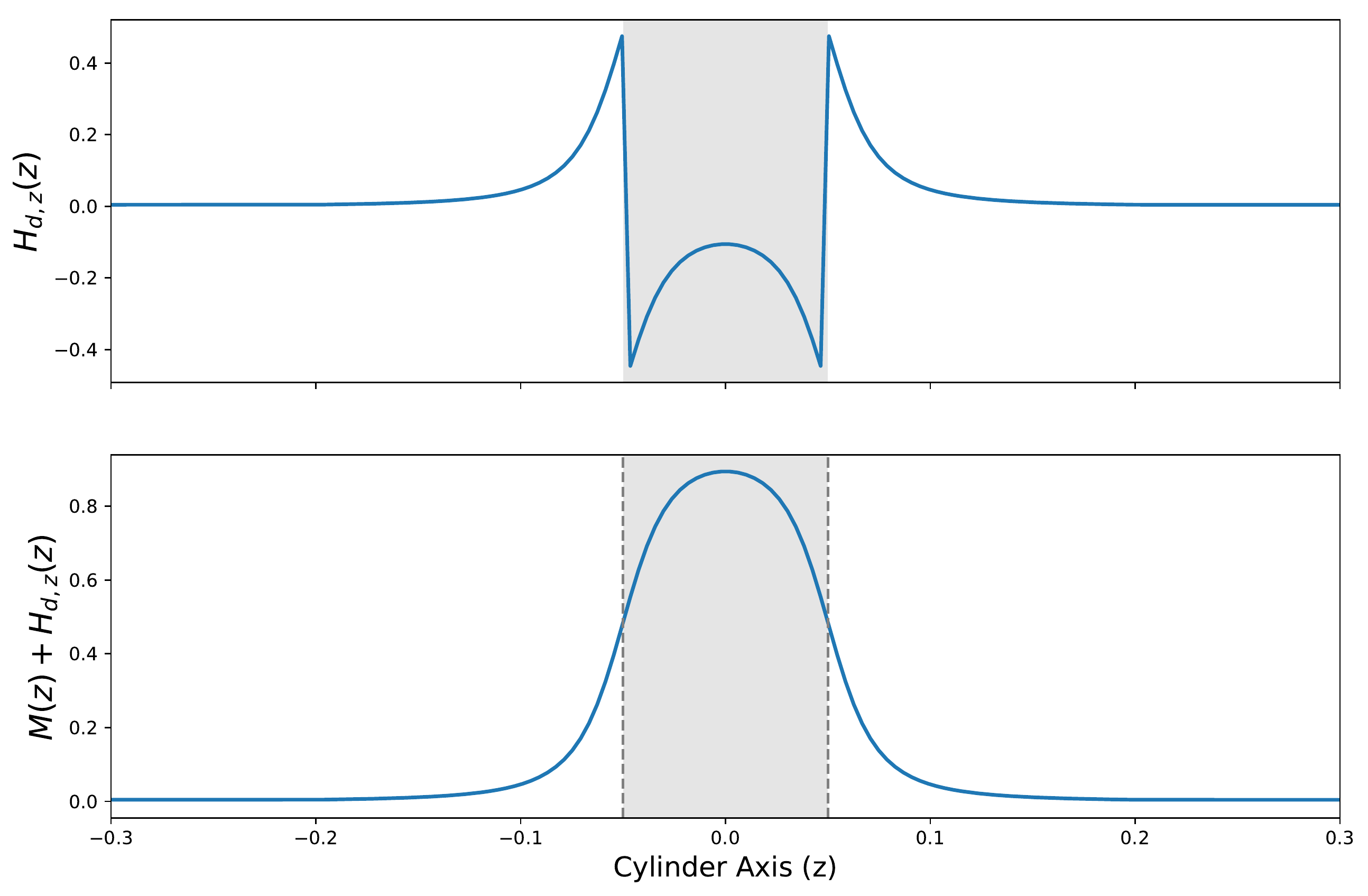}
\end{centering}
\titledcaption[Example of the Demagnetizing Field in a Finite Cylinder]{
The upper plot shows the $N_{zz}$-component
of the demagnetizing tensor in cylinder, based on \cite{Lang2016}.
The lower plot shows the overall effect of a uniformly magnetized
sample and the demagnetizing field. }
\label{demagnetizing-field}
\end{figure}

The analysis so far has ignored the major problem of modeling
magnetostatic modes in a finite cylinder: the demagnetizing field
of the cylinder. For certain simple shapes such as ellipsoids, the demagnetizing field is uniform and proportional to the magnetization, so its contribution is easily handled using demagnetizing factors such as $N_z$:
\begin{equation}
  H_0 \to H = H_0+H_d = H_0-4 \pi N_z M_s.
  \label{applied-minus-demagnetizing-factor}
\end{equation}
Here $H_0$ is the applied field, $H_d$ the demagnetizing field, and $M_s$ the saturation magnetization. There is no such demagnetizing factor for a finite cylinder, as the demagnetizing
field is not uniform (see \cref{demagnetizing-field}), and in fact also has components in the $x$- and $y$-directions. 
A rudimentary attempt to circumvent the problem is using an averaged
demagnetization factor, although as the cylinder is not one of the objects
for which the demagnetizing field is uniform, such an attempt
is at best suspect. Nevertheless one may hope to learn something from
the study of Walker modes by ignoring the complex behavior of the demagnetizing
field. 

The averaged demagnetization factor is given by \cite{joseph-ballistic-factor,demag-elliptic-cylinders}

\begin{align}
  \begin{split}
N_{z}= & 1-\frac{8}{3\pi}\frac{r_{c}}{L}\left[-1+\frac{1}{\varepsilon}\left(\frac{2\varepsilon^{2}-1}{\varepsilon^{2}}E(\varepsilon)+\frac{1-\varepsilon^{2}}{\varepsilon^{2}}K(\varepsilon)\right)\right]\label{eq:demag-factor}\\
\varepsilon^{2}= & \frac{1}{1+\left(\frac{L}{2r_{c}}\right)^{2}}.
  \end{split}
\end{align}

\onecolumngrid

\begin{figure}
\begin{centering}
\includegraphics[width=0.8\textwidth]{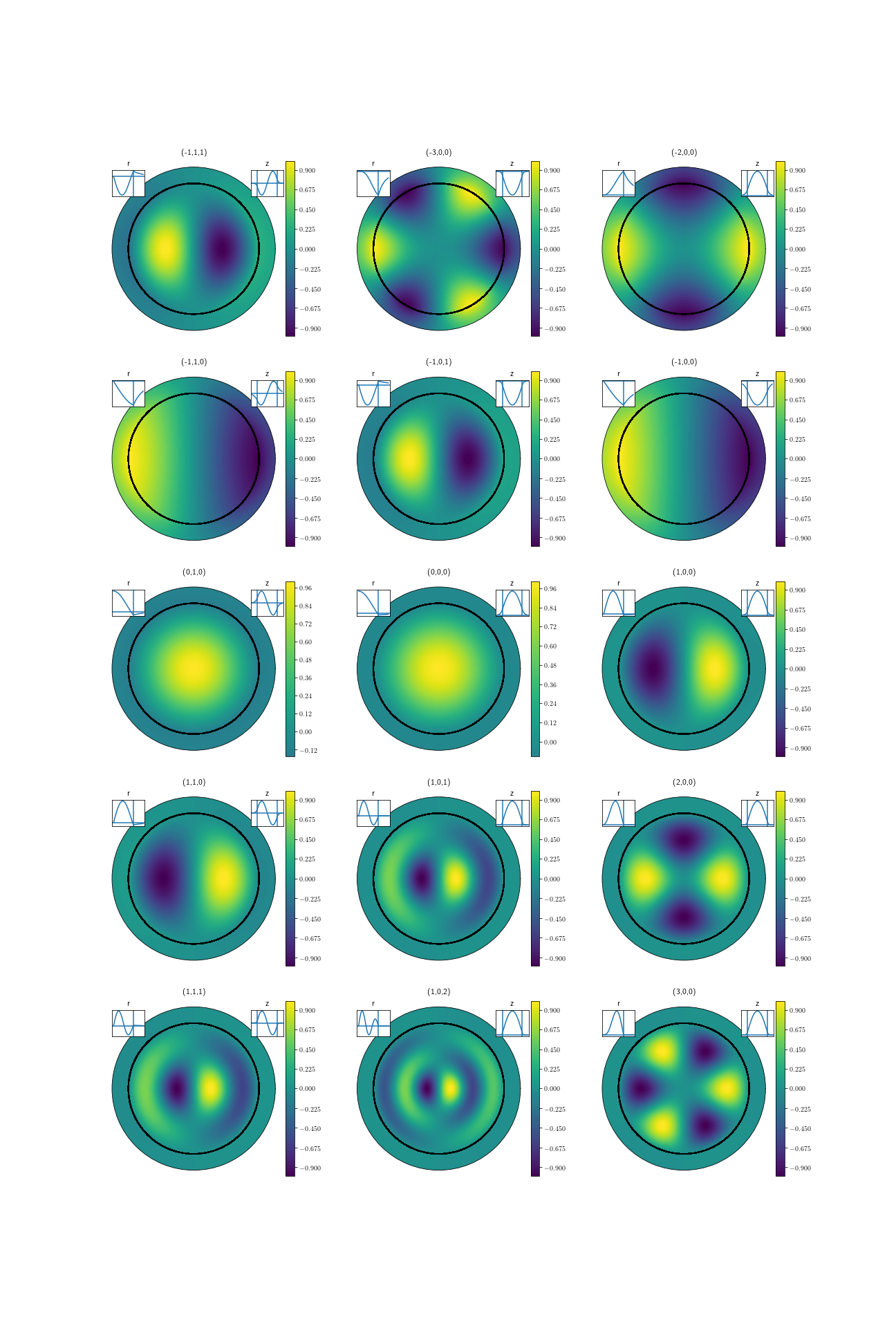}
\end{centering}
\titledcaption[Some examples of mode functions in uniform field]{
The modes are specified by the triple $(l, z_n, r_n)$, where $l$ specifies
the rotation symmetry (azimuthal number), and $z_n$ and $r_n$ are the axial and radial mode numbers, respectively.
}
\label{mode-functions-homogeneous}
\end{figure}

\begin{centering}
\begin{figure}
\includegraphics[width=0.75\textwidth]{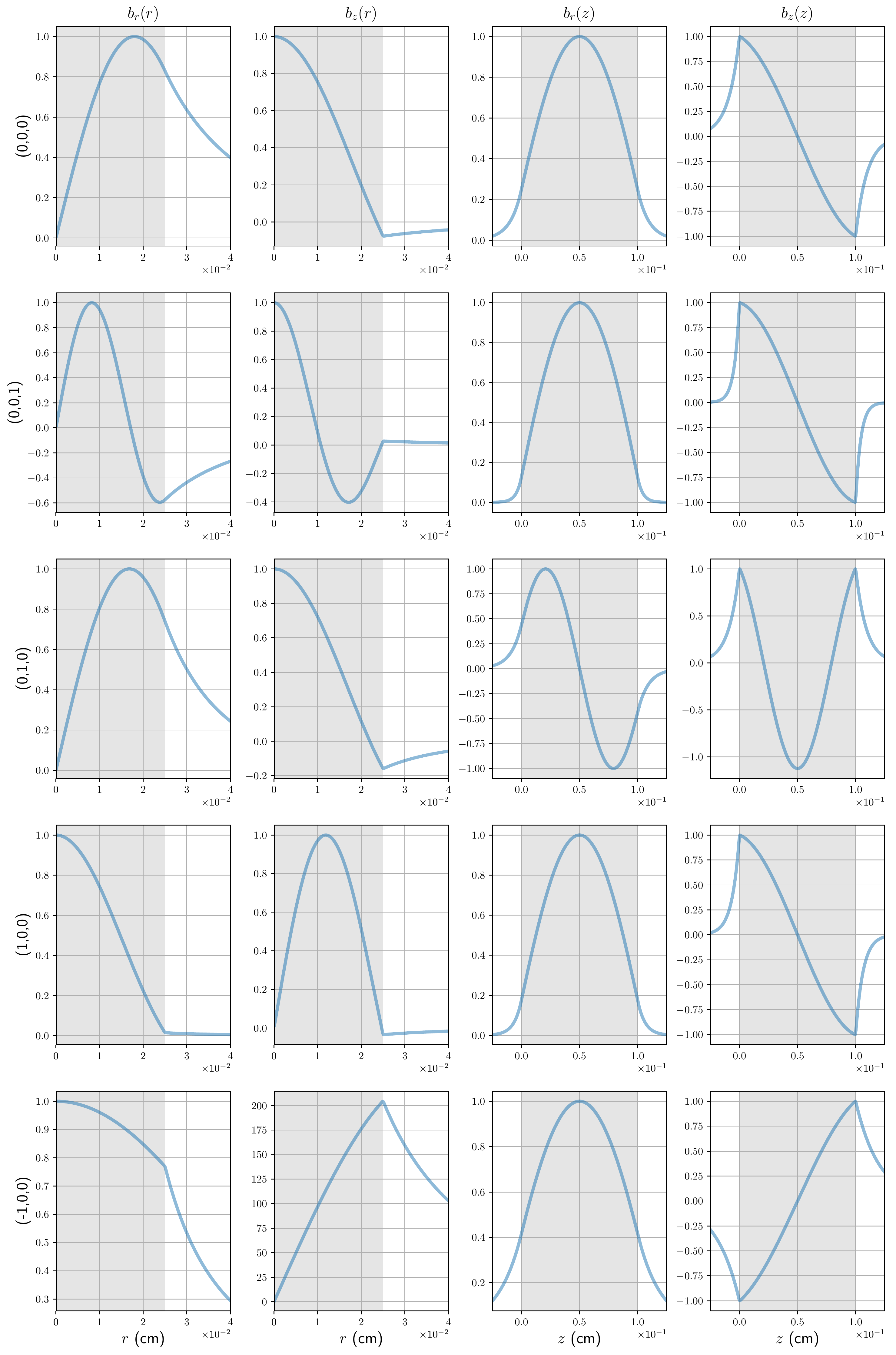}
\titledcaption[{$\protect\vec{b}(\protect\vec{r})$} of select magnetostatic modes in uniform field]{The figure
shows $\vec{b} = \vec{h}+4 \pi \vec{m}$ of select modes; see \cref{mode-functions-homogeneous} for labeling of modes. The shaded area is inside the cylinder.}
\label{brr-functions-homogeneous}
\end{figure}
\end{centering}
\clearpage
\twocolumngrid
\subsection{Absence of Solution for Positive $\mu$}\label[appendix]{positive-mu-absence}
Joseph and Schl{\"o}mann find some solutions also for positive $\mu$ in the infinite cylinder\cite{cylinder}. These modes would be found below the main peak so their absence is highly undesirable. The solution of Joseph and Schl{\"o}mann has an oscillatory nature along the axis of the cylinder, paired with the $I_l$ Bessel-function in the radial direction. In a finite cylinder these modes cannot be glued to a vanishing solution of the Laplace equation at the ends of the cylinder. In contrast, it is entirely possible to write a physical solution with radial $J_l(r)$ and $\sinh(z)$ and $\cosh(z)$, but the resulting dispersion relation has no solution for positive $k$:
\begin{equation}
\tanh\left(k\sqrt{\mu}L\right)=\frac{-2\sqrt{\mu}}{1+\mu}.
\end{equation}
(Solutions with a negative $k$ would not vanish at infinity.) On the other hand, it is easily seen that for different boundary conditions solutions for positive $\mu$ probably exist, for instance for $\Psi(\vec{r})=0$ at the ends. So both the geometry and Maxwell boundary conditions contribute to the lack solutions when $\mu > 0$, as if the field leaking outside would be in some sense problematic. Given that this treatment does not account for the demagnetizing field which does leak out of the cylinder, it is a distinct possibility that modes below the main peak could be found in real physical systems.

\section{Demagnetizing Field in a Finite Cylinder}\label[appendix]{calculating-demagnetizing-field}
The static magnetic field in the cylinder of hydrogen gas has two major components:
the applied (external) magnetic field  $H_0$, and the magnetic field due to the polarized gas, often called the demagnetizing field or stray field. The equation of the demagnetizing field is derived from the Gauss's law for magnetism:
\begin{equation*}
  \nabla\cdot\vec{B}=\nabla \cdot \vec{H} +4\pi\nabla\cdot\vec{M}=0.
\end{equation*}
With $\nabla \times \vec{H}=0$ and $\vec{H} = H_0\vu{e}_z+\vec{H}_d$, $\vec{H}_d=\nabla \Psi$ and we arrive to the equation
\begin{align*}
  \nabla^2 \Psi & = - 4 \pi \nabla \cdot \vec{M} \\
  & = - 4 \pi M_s (\delta(z)-\delta(z-L)).
\end{align*}
This can be solved using the Green's function for a cylinder:
\begin{align*}
  \Psi=& 4\pi M_s \int\frac{\delta\left(z\right)-\delta\left(z-L\right)}{4\pi\sqrt{r^{2}+\left(z-u\right)^{2}}}r\dd{r}\dd{z}\dd{\theta} \\
=& 4\pi M_s\int\left[\frac{1}{2\sqrt{r^{2}+u^{2}}}-\frac{1}{2\sqrt{r^{2}+\left(u-L\right)^{2}}}\right]r\dd{r} \\
  =& -\frac{4\pi M_s}{2} \left(u-\sqrt{r_{c}^{2}+u^{2}}-\sqrt{L^{2}-2 L u+u^{2}}\right. \\
  + & \left. \sqrt{L^{2}-2Lu+r_{c}^{2}+u^{2}}\right)\\
  =& -\frac{4\pi M_s}{2} \left(u-\sqrt{r_{c}^{2}+u^{2}}-\left|L-u\right| \right. \\
  + & \left. \sqrt{L^{2} - 2 L u+r_{c}^{2}+u^{2}} \right) \\
  =& -\frac{4\pi M_s}{2} \left( 2 u-L-\sqrt{r_{c}^{2}+u^{2}} \right. \\
  +& \left. \sqrt{L^{2}- 2 L u+r_{c}^{2}+u^{2}} \right).
\end{align*}

Then $\vec{H_d} = \nabla \Psi$:
\begin{align}
   & H_{d,z}\left(0,\theta,z\right)= \dv{\Psi}{u} \nonumber \\
  =&-\frac{4\pi M_s}{2}\left(2-\frac{u}{\sqrt{r_{c}^{2}+u^{2}}}+\frac{-L+u}{\sqrt{r_{c}^{2}+\left(u-L\right)^{2}}}\right)
  \label{zaxis-demag-field}\\
  \stackrel{u \to L}{\longrightarrow} & -4\pi M_s+\frac{4\pi M_s}{2}\frac{L}{\sqrt{r_{c}^{2}+L^{2}}}
\end{align}
at the ends of the cylinder. As a check, substituting $v=u-\frac{L}{2}$ to \cref{zaxis-demag-field} and taking the
limit of infinite cylinder gives the correct result inside the cylinder:
\begin{align*}
  \vec{B}(z=L) = & \vec{H}+4 \pi \vec{M}  \\
  = &\left(H_0 + H_{d,z}(0, \theta, v) +4 \pi M_s\right) \vu{e}_z \\
  = & \left[H_0+4 \pi M_s - \frac{4\pi M_s}{2}\left(2 - \frac{v+\frac{L}{2}}{\sqrt{r_{c}^{2}+\left(v+\frac{L}{2}\right)^{2}}}\right.\right. \\
  + & \left.\left. \frac{v-\frac{L}{2}}{\sqrt{r_{c}^{2}+\left(v-\frac{L}{2}\right)^{2}}}\right)\right]\vu{e}_z \\
   \stackrel{L\to \infty}{\longrightarrow} &  H_0 \vu{e}_z.
\end{align*}
\section{Solution of the Walker Equation In a Finite Cylinder with a Linearly
  Increasing Magnetic Field}\label[appendix]{linear-gradient-solution}
 In a sense the presence of a linear gradient simplifies
the treatment as the static case may now be considered solved, i.e.
\[
\vec{H}\left(z\right)=H_{0}\vu{e}_{z}+\vec{H}_{1}\left(\vec{r}\right)+4\pi\vec{M}\left(\vec{r}\right)+\vec{H}_{d}\left(\vec{r}\right)\approx\left[H_{0}+Cz\right]\vu{e}_{z},
\]
where $\vec{H}_{1}\left(\vec{r}\right)$ denotes other magnetic fields
present in the setup, and $\vec{H}_{d}\left(\vec{r}\right)$ is the
demagnetizing field. This suggests an approach where $H_{0}$ is just replaced
with $H_{0}+Cz$ in the Walker equation; it turns out this is indeed the case,
but not trivially so. The substitution makes $\kappa,\nu,$ and $\mu$
into functions of $z$. Generally one must be careful with such substitutions
and rederive the Walker equation (or use the generalized Walker equation \cite{Arias2015});
in this case, the $z$-dependence of $\kappa$ and $\nu$ plays no
role in \cref{eq:walker-derivation} and we have

\[
\kappa\left(z\right)=\frac{4\pi\gamma^{2}\left(H_{0}+Cz\right)M_{0}}{\gamma^{2}\left(H_{0}+Cz\right)^{2}-\omega^{2}}.
\]

However, a modification occurs when the Gauss law $\nabla\cdot\vec{B}$
is applied, and the resulting Walker equation inside the cylinder
turns out to be

\begin{equation}
\left[1+\kappa\left(z\right)\right]\left[\pdv[2]{r}+\frac{1}{r}\pdv{r}+\frac{1}{r^{2}}\pdv[2]{\theta}\right]\Phi+\pdv[2]{\Phi}{z}+C=0.\label{eq:linear-walker-eq}
\end{equation}

The equation has a $z$-dependent coefficient with potentially troublesome
behavior for some values of $\omega$ and $z$, and the constant
$C$ makes it not separable; had we not rederived the Walker equation
but just modified $\kappa$, we would have missed the latter. Fortunately,
it turns out the separability is not a problem: $\psi\left(z\right)=-\frac{C}{2}z^{2}$
is a solution of the equation, so substituting $\Phi\left(\vec{r}\right)=\Psi\left(r,z\right)$+$\psi\left(z\right)$
gets rid of the constant and reduces \cref{eq:linear-walker-eq} to
a separable Walker equation for $\Phi\left(\vec{r}\right)$. $\psi\left(z\right)$
is also a solution of the Laplace equation, so that part of the solution
inside and outside the cylinder at the sides can be matched, supposing
the magnetic field also has a similar gradient outside the cylinder.
At the ends of the cylinder it is perhaps easiest to assume that the
magnetic field attains its maximum or minimum value and remains constant
outside the cylinder: this way the continuity of the magnetic field
cancels delta-function contributions coming from $\partial_{z}\left(Cz\right)$
at the sides of the cylinder. The continuity of $\Phi$ across the
ends requires also that the solution outside include $\frac{CL^{2}}{2}$
(also a solution of Laplace equation) at the $z=L$ end.

$\kappa$ and $\nu$ being functions of $z$ significantly complicates
solving the boundary conditions. For one, it is not trivial to match
the $z$-dependent solution of the Laplace equation and the Walker
equation at the sides of the cylinder; for another, \cref{eq:radial-boundary-condition}
has also become $z$-dependent. To facilitate finding a solution,
we modify the boundary conditions. For the radial boundary condition,
it is easiest to choose either the vanishing of the radial function
or its derivative at the boundary; this should not matter much as
the equation is still separable and we are not really interested in
changing the radius of the cylinder which is fixed in our experiments.
In principle there is no need to modify the axial boundary conditions,
but for ease we have done so. While this may affect the resulting mode
behavior we hope something can be learnt from it. We have chosen the
somewhat natural $\dv{\Psi}{z}=h_{z}=0$ at the ends of the cylinder.

The following redefinitions of the parameters are used:
\begin{align*}
\Omega_{H}= & \frac{4\pi M_{s}}{H_{0}},\alpha=\frac{C}{H_{0}},\tilde{\omega}=\frac{\omega}{\gamma H_{0}},\\
\kappa\left(z\right)= & \frac{\Omega_{H}\left(1-\alpha z\right)}{\left(1-\alpha z\right)^{2}-\tilde{\omega}^{2}}.
\end{align*}

The separable Walker equation can be solved with $\Psi\left(r,\theta,z\right)=$$Z\left(z\right)\phi\left(r,\theta\right)$. The radial equation just gives the Bessel equation whose solutions
inside the cylinder must be $J_{l}\left(kr\right)$, and the azimuthal
solutions are the usual $e^{il\theta}$. We immediately choose $l=0$. The axial equation becomes
\begin{align*}
 k^{2}= &\frac{\dv[2]{Z}{z}}{Z\left(z\right)\left(1+\kappa\left(z\right)\right)} \Leftrightarrow\\
  \dv[2]{Z}{z}= & \frac{k^{2}Z\left(z\right)}{\tilde{\omega}^{2}-\left(2-\alpha z\right)^{2}} \\
  \times & \left[\tilde{\omega}^{2}-\left(1-\alpha z\right)^{2}-\Omega_{H}\left(1-\alpha z\right)\right].
\end{align*}

Substituting $y=\alpha z+\tilde{\omega}-1$ gives 
\begin{align*}
\alpha^{2}\dv[2]{Z}{y}= & \frac{k^{2}Z\left(y\right)}{y\left(2\tilde{\omega}-y\right)}\left[y\left(2\tilde{\omega}-y\right)+\Omega_{H}y-\Omega_{H}\tilde{\omega}\right]
\end{align*}

whose partial fraction decomposition is
\begin{equation}
\alpha^{2}\dv[2]{Z}{z}=k^{2}Z\left(y\right)\left[1-\frac{\Omega_{H}}{2y}+\frac{\Omega_{H}}{2\left(2\tilde{\omega}-y\right)}\right].\label{eq:unapproximated-gradient-diffeq}
\end{equation}

The experimentally observed frequencies are small compared to the
static field, so that typically $\omega\approx$$\gamma H_{0}\left(1+\delta\right)$
with $\delta\sim10^{-5}-10^{-6}$. With $\alpha\sim\SI{1e-6}{\per\cm}$and
$z\sim\SI{0.1}{\cm}$, $y\approx 10^{-7}+1+\delta-1\approx\delta\leq10^{-4}$.
$2\tilde{\omega}-y$ is just $\tilde{\omega}-\alpha z+1$ which (for
positive frequencies) is close to 2. Hence we neglect the
last term of \cref{eq:unapproximated-gradient-diffeq} as small by
at least 4 orders of magnitude: 
\begin{align*}
\alpha^{2}\dv[2]{Z}{y}= & k^{2}Z\left(y\right)\left[1-\frac{\Omega_{H}}{2y}\right].
\end{align*}

This equation has a solution in terms of the confluent hypergeometric
function and the Kummer/Tricomi $U$ function (with $\bar{\omega}=\tilde{\omega}-1$):
\begin{align*}
Z\left(z\right)= & \left[A\;\pFq{1}{1}{1-\frac{\Omega_{H}k}{4\alpha}}{2}{\frac{2k}{\alpha}\left(\alpha z+\bar{\omega}\right)}\right.\\
  + & \left.B\,\mathrm{U}\left(1-\frac{\Omega_{H}k}{4\alpha},2,\frac{2k}{\alpha}\left(\alpha z+\bar{\omega}\right)\right)\right] \\
  \times & \left(\alpha z+\bar{\omega}\right) e^{-\frac{k}{\alpha}\left(\alpha z+\bar{\omega}\right)}.
\end{align*}
The frequencies $\bar{\omega}$ are given by imposing the boundary
conditions at the ends of the cylinder. The resulting two equations
(one at $z=0$ and the other $z=L$) can be combined into one equation
whose roots give the $\bar{\omega}$:

\begin{align}\label{linear-gradient-characteristic-equation}
  \begin{split}
  0 = & \frac{F_{0}^{L}k}{2}\left(-4U_{0}^{0}\alpha\left(\alpha-k\overline{\omega}\right)
  +2U_{1}^{0}k\overline{\omega}\left(4\alpha-\Omega_{H}k\right)\right) \\
  \times & \left(L\Omega_{H}\alpha k-4L\alpha^{2}+\Omega_{H}k\overline{\omega}-4\alpha\overline{\omega}\right)\\
  +& 2F_{1}^{L}\alpha\left(-4U_{0}^{0}\alpha\left(\alpha-k\overline{\omega}\right)
  +2U_{1}^{0}k\overline{\omega}\left(4\alpha-\Omega_{H}k\right)\right) \\
  \times & \left(L\alpha k-\alpha+k\overline{\omega}\right)\\
 - & U_{0}^{L}k\left(4F_{0}^{0}\alpha\left(\alpha-k\overline{\omega}\right)
  +F_{1}^{0}k\overline{\omega}\left(4\alpha-\Omega_{H}k\right)\right) \\
  \times & \left(L\Omega_{H}\alpha k-4L\alpha^{2}+\Omega_{H}k\overline{\omega}-4\alpha\overline{\omega}\right)\\
  +& 2U_{1}^{L}\alpha\left(4F_{0}^{0}\alpha\left(\alpha-k\overline{\omega}\right)
  +F_{1}^{0}k\overline{\omega}\left(4\alpha-\Omega_{H}k\right)\right)\\
\times & \left(L\alpha k-\alpha+k\overline{\omega}\right) \\
F_{i}^{t}\coloneqq & \pFq{1}{1}{1+i-\frac{\Omega_{H}k}{4\alpha}}{2+i}{\frac{2k}{\alpha}\left(\alpha t+\bar{\omega}\right)}\\
U_{i}^{t}\coloneqq & \mathrm{U}\left(1+i-\frac{\Omega_{H}k}{4\alpha},2+i,\frac{2k}{\alpha}\left(\alpha t+\bar{\omega}\right)\right).
\end{split}
\end{align}
The equations were derived in a Jupyter notebook \cite{jupyter} using the SymPy package \cite{sympy}; finding the numerical solutions required at times arbitrary precision arithmetic and special functions of the mpmath package \cite{mpmath}. Figures were produced with matplotlib \cite{matplotlib}.

\begin{figure*}[b]
  \centering
  \begin{minipage}[b]{0.45\textwidth}
  \includegraphics[width=\textwidth]{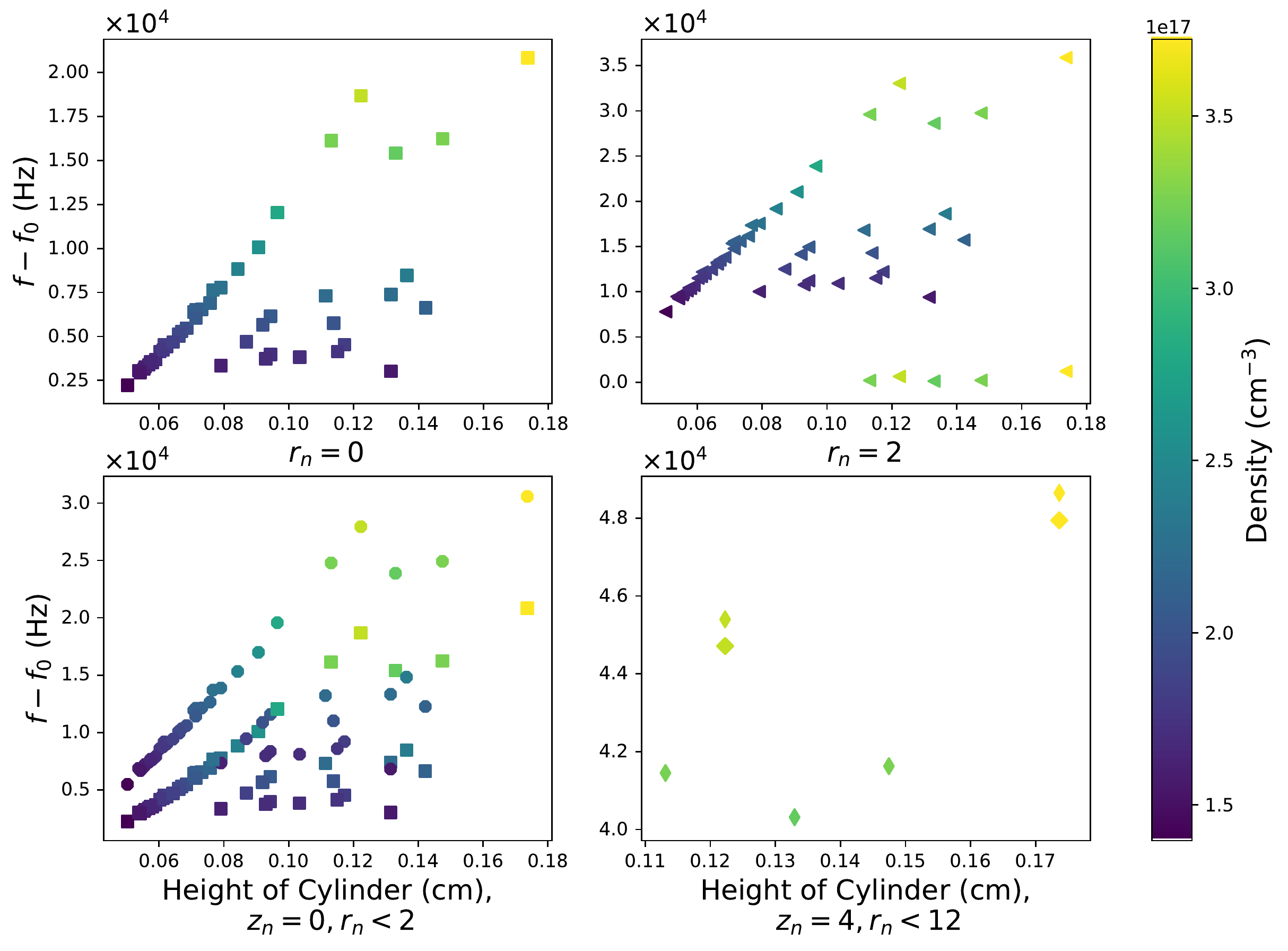}
\titledcaption[Distribution of modes in linear gradient]{
$\alpha=6.52\cdot10^{-5}$, corresponding to gradient $C=3.0\text{G/cm}$. For convenience,
$f - \frac{\gamma H_0}{2 \pi}$ has been plotted. Only one axial mode seems to correspond to the
first radial mode. For higher radial modes (different markers) some more axial modes are present, albeit
not all that many $z_n=4$ modes were found for the first 12 radial modes. 
} \label{linear-gradient-3.0-distribution-of-modes}
  \end{minipage}\hfill
  \begin{minipage}[b]{0.45\textwidth}
\includegraphics[width=\textwidth]{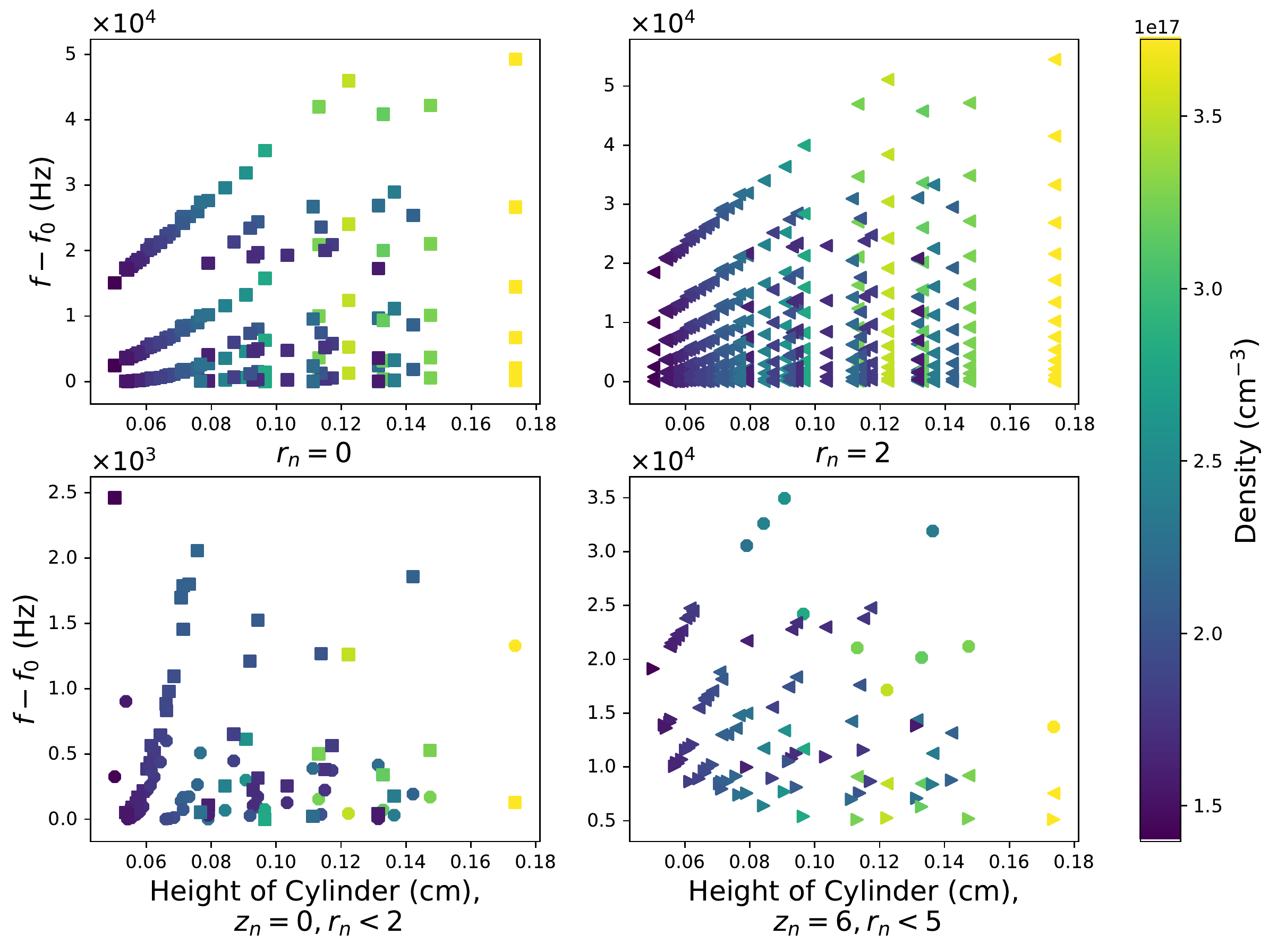}
\titledcaption[Distribution of Modes in Linear Gradient]{
$\alpha=6.52\cdot10^{-6}$, corresponding to gradient $C=0.3\text{G/cm}$. For convenience,
$f - \frac{\gamma H_0}{2 \pi}$ has been plotted. The number of modes is clearly greater than in the case of a higher gradient. }
\label{linear-gradient-0.3-distribution-of-modes}
  \end{minipage}
\end{figure*}

\Cref{linear-gradient-3.0-distribution-of-modes} shows how the modes are distributed as function of experimental data. For low radial modes numbers, only one axial mode is present, though the number of axial modes
increases slowly with the radial mode number for these parameters.
For a lower gradient (\cref{linear-gradient-0.3-distribution-of-modes}),
more axial modes are found for even the first radial mode,
and the number clearly increases for higher modes. The number of axial
modes seems to increase with higher radial mode numbers, although not that many $z_n = 4$ modes were found among the first 12 radial modes; those that were found occurred at relatively high densities. This seems to support the idea that the gradient reduces the magnetostatic character of the modes, i.e. the number of magnetostatic modes decreases as the gradient increases, as is readily seen from \cref{linear-gradient-0.3-distribution-of-modes}.
\end{document}

%% file: cylinder-area.tex
\tikzset{
	MyPersp/.style={scale=1.5,x={(-0.8cm,-0.4cm)},y={(0.8cm,-0.4cm)},
    z={(0cm,1cm)}},
	MyPoints/.style={fill=white,draw=black,thick}
		}
\begin{tikzpicture}[MyPersp,font=\large]
	\def\h{2.5}
	\def\hh{2}
	\def\a{35}
	\def\aa{35}
	\def\thetaa{315}
	\def\thetaaa{135}
	\def\r1{2}

	\fill[gray!20, opacity=0.8] ({\r1*cos(\thetaa)},{\r1*sin(\thetaa)},0) -- ++(0,0,0) arc[start angle=\thetaa,delta angle=180,radius=\r1] -- ++(0,0,{-\hh}) arc[start angle=\thetaa+180,delta angle=-180,radius=\r1];

	\fill[red!10, opacity=1] ({cos(\thetaa)},{sin(\thetaa)},0) -- ++(0,0,0) arc[start angle=\thetaa,delta angle=180,radius=1] -- ++(0,0,{-\hh}) arc[start angle=\thetaa+180,delta angle=-180,radius=1];
	\fill[green!10, opacity=0.8] ({\r1*cos(\thetaa)},{\r1*sin(\thetaa)},0) -- ++(0,0,0) arc[start angle=\thetaa,delta angle=180,radius=\r1] -- ++(0,0,{2.0*\h}) arc[start angle=\thetaa+180,delta angle=-180,radius=\r1];
	\fill[blue!10, opacity=0.8] ({cos(\thetaa)},{sin(\thetaa)},0) -- ++(0,0,0) arc[start angle=\thetaa,delta angle=180,radius=1] -- ++(0,0,{2.0*\h}) arc[start angle=\thetaa+180,delta angle=-180,radius=1];
	\fill[gray!20, opacity=0.8] ({\r1*cos(\thetaa)},{\r1*sin(\thetaa)},{2.0*\h}) -- ++(0,0,0) arc[start angle=\thetaa,delta angle=180,radius=\r1] -- ++(0,0,\hh) arc[start angle=\thetaa+180,delta angle=-180,radius=\r1];
	\fill[gray!20, opacity=0.8] ({\r1*cos(\thetaaa)},{\r1*sin(\thetaaa)},{2.0*\h}) -- ++(0,0,0) arc[start angle=\thetaaa,delta angle=180,radius=\r1] -- ++(0,0,\hh) arc[start angle=\thetaaa+180,delta angle=-180,radius=\r1];
	\fill[red!10, opacity=1] ({cos(\thetaaa)},{sin(\thetaaa)},{2.0*\h}) -- ++(0,0,0) arc[start angle=\thetaaa,delta angle=180,radius=1] -- ++(0,0,\hh) arc[start angle=\thetaaa+180,delta angle=-180,radius=1];
	\fill[red!10, opacity=1] ({cos(\thetaa)},{sin(\thetaa)},{2.0*\h}) -- ++(0,0,0) arc[start angle=\thetaa,delta angle=180,radius=1] -- ++(0,0,\hh) arc[start angle=\thetaa+180,delta angle=-180,radius=1];

	\node [anchor=west, font=\huge] at ({\r1*cos(\thetaaa)},{\r1*sin(\thetaaa)},0) {$z=0$};
	\node [anchor=west, font=\huge] at ({\r1*cos(\thetaaa)},{\r1*sin(\thetaaa)},{2*\h}) {$z=L$};
	\node [font=\huge] at (0,0,{\h}) {A};
	\node [font=\huge] at ({(\r1+1)*0.5*cos(\thetaa)},{(\r1+1)*0.5*sin(\thetaa)},{\h}) {B};
	\node [font=\huge] at (0,0,{2*\h+\hh}) {C};	
	\node [font=\huge] at (0,0,{-\hh}) {C};	
	\node [font=\huge] at ({(\r1+1)*0.5*cos(\thetaa)},{(\r1+1)*0.5*sin(\thetaa)},{2.0*\h+\hh})
	 {D};
	 \node [font=\huge] at ({(\r1+1)*0.5*cos(\thetaa)},{(\r1+1)*0.5*sin(\thetaa)},{-\hh}) {D};
	\draw (0,0,0)  -- ({cos(\thetaaa)},{sin(\thetaaa)},0);
	\node [font=\huge] at ({0.3*cos(\thetaaa)},{0.3*sin(\thetaaa)},0.2)  {$r_c$};

	\foreach \t in {\thetaa, \thetaaa}
		\draw[black,thick] ({cos(\t)},{sin(\t)},0)
      --({cos(\t)},{sin(\t)},{2.0*\h});
	\draw[black,very thick] (1,0,0) 
		\foreach \t in {5,10,...,360}
			{--({cos(\t)},{sin(\t)},0)}--cycle;
	\draw[black,very thick] (1,0,{2*\h}) 
		\foreach \t in {10,20,...,360}
			{--({cos(\t)},{sin(\t)},{2*\h})}--cycle;

	\foreach \t in {\thetaa, \thetaaa}
		\draw[black,thick] ({cos(\t)},{sin(\t)},{2.0*\h})
      --({cos(\t)},{sin(\t)},{2.0*\h+\hh});
	\draw[black,very thick] (1,0,{2.0*\h}) 
		\foreach \t in {5,10,...,360}
			{--({cos(\t)},{sin(\t)},{2.0*\h})}--cycle;
	\draw[black,dashed] (1,0,{2*\h+\hh}) 
		\foreach \t in {10,20,...,360}
			{--({cos(\t)},{sin(\t)},{2*\h+\hh})}--cycle;
	
	\foreach \t in {\thetaa, \thetaaa}
		\draw[black,dashed] ({\r1*cos(\t)},{\r1*sin(\t)},0)
      --({\r1*cos(\t)},{\r1*sin(\t)},{2.0*\h});
	\draw[black,very thick] ({\r1},0,0) 
		\foreach \t in {5,10,...,360}
			{--({\r1*cos(\t)},{\r1*sin(\t)},0)}--cycle;
	\draw[black,very thick] ({\r1},0,{2*\h}) 
		\foreach \t in {10,20,...,360}
			{--({\r1*cos(\t)},{\r1*sin(\t)},{2*\h})}--cycle;

	\foreach \t in {\thetaa, \thetaaa}
		\draw[black,dashed] ({\r1*cos(\t)},{\r1*sin(\t)},{2*\h})
      --({\r1*cos(\t)},{\r1*sin(\t)},{2.0*\h+\hh});
	\draw[black,very thick] ({\r1},0,{2*\h}) 
		\foreach \t in {5,10,...,360}
			{--({\r1*cos(\t)},{\r1*sin(\t)},{2*\h})}--cycle;
	\draw[black, dashed] ({\r1},0,{2*\h+\hh}) 
		\foreach \t in {10,20,...,360}
			{--({\r1*cos(\t)},{\r1*sin(\t)},{2*\h+\hh})}--cycle;

	\foreach \t in {\thetaa, \thetaaa}
		\draw[black,thick] ({cos(\t)},{sin(\t)},{0})
      --({cos(\t)},{sin(\t)},{-\hh});
	\draw[black,very thick] (1,0,0) 
		\foreach \t in {5,10,...,360}
			{--({cos(\t)},{sin(\t)},0)}--cycle;
	\draw[black,dashed] (1,0,{-\hh}) 
		\foreach \t in {10,20,...,360}
			{--({cos(\t)},{sin(\t)},{-\hh})}--cycle;

	\foreach \t in {\thetaa, \thetaaa}
		\draw[black,dashed] ({\r1*cos(\t)},{\r1*sin(\t)},0)
      --({\r1*cos(\t)},{\r1*sin(\t)},{-\hh});
	\draw[black,very thick] ({\r1},0,0) 
		\foreach \t in {5,10,...,360}
			{--({\r1*cos(\t)},{\r1*sin(\t)},0)}--cycle;
	\draw[black, dashed] ({\r1},0,{-\hh}) 
		\foreach \t in {10,20,...,360}
			{--({\r1*cos(\t)},{\r1*sin(\t)},{-\hh})}--cycle;
\end{tikzpicture}